\documentclass[3p]{elsarticle}

\usepackage{lineno}

\usepackage{amsmath}
\usepackage{amssymb}
\usepackage{subfigure}
\usepackage{epsfig}
\usepackage{siunitx}
\DeclareSIUnit\Molar{M}     %

\usepackage{color}

\usepackage{hyperref}   %

\makeatletter %

\def\input@path{{01_sections/}{02_figures/}}

\newcommand{\secref}[1]{Section~\ref{#1}}%
\newcommand{\figref}[1]{Figure~\ref{#1}}%

\newcommand\DeclareBoldMathCommand[2]{%
  \protected@edef\@tempb{%
    \noexpand\DeclareRobustCommand{\csname #1\endcsname}{\boldsymbol{\ensuremath{#2}}}}
  \@tempb}
\@tfor\AM@letter:=abcdefghijklmnopqrstuvwxyz%
\do{\DeclareBoldMathCommand{v\AM@letter}{\AM@letter}}
\@tfor\AM@letter:=ABCDEFGHIJKLMNOPQRSTUVWXYZ%
\do{\DeclareBoldMathCommand{v\AM@letter}{\AM@letter}}
\DeclareBoldMathCommand{vnull}{0}
\DeclareBoldMathCommand{vone}{1}
\DeclareBoldMathCommand{valpha}{\alpha}
\DeclareBoldMathCommand{vbeta}{\beta}
\DeclareBoldMathCommand{vgamma}{\gamma}
\DeclareBoldMathCommand{vdelta}{\delta}
\DeclareBoldMathCommand{vepsilon}{\epsilon}
\DeclareBoldMathCommand{vvarepsilon}{\varepsilon}
\DeclareBoldMathCommand{vzeta}{\zeta}
\DeclareBoldMathCommand{veta}{\eta}
\DeclareBoldMathCommand{vtheta}{\theta}
\DeclareBoldMathCommand{vvartheta}{\vartheta}
\DeclareBoldMathCommand{viota}{\iota}
\DeclareBoldMathCommand{vkappa}{\kappa}
\DeclareBoldMathCommand{vlambda}{\lambda}
\DeclareBoldMathCommand{vmu}{\mu}
\DeclareBoldMathCommand{vnu}{\nu}
\DeclareBoldMathCommand{vpi}{\pi}
\DeclareBoldMathCommand{vxi}{\xi}
\DeclareBoldMathCommand{vrho}{\varrho}
\DeclareBoldMathCommand{vsigma}{\sigma}
\DeclareBoldMathCommand{vtau}{\tau}
\DeclareBoldMathCommand{vupsilon}{\upsilon}
\DeclareBoldMathCommand{vphi}{\varphi}
\DeclareBoldMathCommand{vchi}{\chi}
\DeclareBoldMathCommand{vpsi}{\psi}
\DeclareBoldMathCommand{vomega}{\omega}
\DeclareBoldMathCommand{vGamma}{\Gamma}
\DeclareBoldMathCommand{vDelta}{\Delta}
\DeclareBoldMathCommand{vTheta}{\Theta}
\DeclareBoldMathCommand{vLambda}{\Lambda}
\DeclareBoldMathCommand{vXi}{\Xi}
\DeclareBoldMathCommand{vPi}{\Pi}
\DeclareBoldMathCommand{vSigma}{\Sigma}
\DeclareBoldMathCommand{vUpsilon}{\Upsilon}
\DeclareBoldMathCommand{vPhi}{\Phi}
\DeclareBoldMathCommand{vPsi}{\Psi}
\DeclareBoldMathCommand{vOmega}{\Omega}

\newcommand\DeclareDiscreteBoldMathCommand[2]{%
  \protected@edef\@tempc{%
    \noexpand\DeclareRobustCommand{\csname #1\endcsname}{\boldsymbol{\mathrm{#2}}}}
  \@tempc}
\@tfor\AM@letter:=abcdefghijklmnopqrstuvwxyz%
\do{\DeclareDiscreteBoldMathCommand{vd\AM@letter}{\AM@letter}}
\@tfor\AM@letter:=ABCDEFGHIJKLMNOPQRSTUVWXYZ%
\do{\DeclareDiscreteBoldMathCommand{vd\AM@letter}{\AM@letter}}
\DeclareDiscreteBoldMathCommand{vdnull}{0}
\DeclareDiscreteBoldMathCommand{vdone}{1}
\DeclareDiscreteBoldMathCommand{vdalpha}{\alpha}
\DeclareDiscreteBoldMathCommand{vdbeta}{\beta}
\DeclareDiscreteBoldMathCommand{vdgamma}{\gamma}
\DeclareDiscreteBoldMathCommand{vddelta}{\delta}
\DeclareDiscreteBoldMathCommand{vdepsilon}{\epsilon}
\DeclareDiscreteBoldMathCommand{vdvarepsilon}{\varepsilon}
\DeclareDiscreteBoldMathCommand{vdzeta}{\zeta}
\DeclareDiscreteBoldMathCommand{vdeta}{\eta}
\DeclareDiscreteBoldMathCommand{vdtheta}{\theta}
\DeclareDiscreteBoldMathCommand{vdvartheta}{\vartheta}
\DeclareDiscreteBoldMathCommand{vdiota}{\iota}
\DeclareDiscreteBoldMathCommand{vdkappa}{\kappa}
\DeclareDiscreteBoldMathCommand{vdlambda}{\lambda}
\DeclareDiscreteBoldMathCommand{vdmu}{\mu}
\DeclareDiscreteBoldMathCommand{vdnu}{\nu}
\DeclareDiscreteBoldMathCommand{vdpi}{\pi}
\DeclareDiscreteBoldMathCommand{vdxi}{\xi}
\DeclareDiscreteBoldMathCommand{vdrho}{\varrho}
\DeclareDiscreteBoldMathCommand{vdsigma}{\sigma}
\DeclareDiscreteBoldMathCommand{vdtau}{\tau}
\DeclareDiscreteBoldMathCommand{vdupsilon}{\upsilon}
\DeclareDiscreteBoldMathCommand{vdphi}{\varphi}
\DeclareDiscreteBoldMathCommand{vdchi}{\chi}
\DeclareDiscreteBoldMathCommand{vdpsi}{\psi}
\DeclareDiscreteBoldMathCommand{vdomega}{\omega}
\DeclareDiscreteBoldMathCommand{vdGamma}{\Gamma}
\DeclareDiscreteBoldMathCommand{vdDelta}{\Delta}
\DeclareDiscreteBoldMathCommand{vdTheta}{\Theta}
\DeclareDiscreteBoldMathCommand{vdLambda}{\Lambda}
\DeclareDiscreteBoldMathCommand{vdXi}{\Xi}
\DeclareDiscreteBoldMathCommand{vdPi}{\Pi}
\DeclareDiscreteBoldMathCommand{vdSigma}{\Sigma}
\DeclareDiscreteBoldMathCommand{vdUpsilon}{\Upsilon}
\DeclareDiscreteBoldMathCommand{vdPhi}{\Phi}
\DeclareDiscreteBoldMathCommand{vdPsi}{\Psi}
\DeclareDiscreteBoldMathCommand{vdOmega}{\Omega}

\providecommand*{\dd}{%
  \@ifnextchar^{\@dd}{\@dd^{}}}
\def\@dd^#1{%
  \mathop{\mathrm{\mathstrut d}}%
  \nolimits^{#1}\dd@gobblespace}
\def\dd@gobblespace{%
  \futurelet\diffarg\dd@opspace}
\def\dd@opspace{%
  \let\dd@space\!%
  \ifx\diffarg(
\let\dd@space\relax%
\else%
\ifx\diffarg[
\let\dd@space\relax%
\else%
\ifx\diffarg\{%
\let\dd@space\relax%
\fi%
\fi%
\fi%
\dd@space}

\newcommand{\Frac}{%
  \@ifnextchar[
  {\Frac@i}
  {\Frac@ii}}
\newcommand{\Frac@i}{}
\def\Frac@i[#1]#2#3{%
  \genfrac{}{}{#1}{}{\displaystyle{#2}}{\displaystyle{#3}}}
\newcommand{\Frac@ii}[2]{\frac{\displaystyle{#1}}{\displaystyle{#2}}}
      \newcommand{\diff@diffspace}{\,}
\newcommand{\diff@mathfrac}[2]{\frac{#1}{#2}}
\newcommand{\diff@mathFrac}[2]{\Frac{#1}{#2}}
\newcommand{\diff@textfrac}[2]{%
  \bgroup #1\egroup\mkern-1mu/\mkern-1mu\bgroup #2\egroup}
\newcommand{\diff}{%
  \global\let\diff@diffop\dd
  \global\let\diff@frac\diff@mathfrac
  \@ifnextchar[
  {\diff@i}
  {\diff@ii}}
\newcommand{\Diff}{%
  \global\let\diff@diffop\dd
  \global\let\diff@frac\diff@mathFrac
  \@ifnextchar[
  {\diff@i}
  {\diff@ii}}
\newcommand{\tdiff}{%
  \global\let\diff@diffop\dd
  \global\let\diff@frac\diff@textfrac
  \@ifnextchar[
  {\diff@i}
  {\diff@ii}}
\newcommand{\pdiff}{%
  \global\let\diff@diffop\partial
  \global\let\diff@frac\diff@mathfrac
  \@ifnextchar[
  {\diff@i}
  {\diff@ii}}
\newcommand{\Pdiff}{%
  \global\let\diff@diffop\partial
  \global\let\diff@frac\diff@mathFrac
  \@ifnextchar[
  {\diff@i}
  {\diff@ii}}
\newcommand{\tpdiff}{%
  \global\let\diff@diffop\partial
  \global\let\diff@frac\diff@textfrac
  \@ifnextchar[
  {\diff@i}
  {\diff@ii}}

\newcommand*{\diff@i}{}
\def\diff@i[#1]#2#3{\eval{\diff@ii{#2}{#3}}_{#1}}

\newcommand*{\diff@ii}[2]{%
  \begingroup
  \toks0={}\count0=0
  \diff@degree #2\diff@degree
  \diff@frac{\diff@diffop\ifnum\count0>1^{\the\count0}\fi\diff@diffspace#1}%
  {\the\toks0}%
  \endgroup}

\newcommand*{\diff@degree}[1]{%
  \ifx #1\diff@degree \expandafter\diff@stopd
  \else \expandafter\diff@addd \fi #1^1$#1\diff@addd}
\newcommand{\diff@stopd}{}
\def\diff@stopd #1\diff@addd{}
\newcommand*{\diff@addd}{}
\def\diff@addd #1^#2#3$#4\diff@addd{%
  \advance\count0 #2
  \toks0=\expandafter{\the\toks0%
    {\diff@diffop\diff@diffspace #4}%
    \diff@diffspace}\diff@degree}

\def\rs#1{\@ifnextchar[
  {\@rs{#1}}{\@@rs{#1}}}
\def\@rs#1[#2]#3{\mathinner{%
    \setbox\@ne\hbox{$\displaystyle{\vphantom{#3}}#1{#3}\m@th$}%
    \setbox\tw@\hbox{$\displaystyle{#3}#2\m@th$}%
    \hskip\wd\@ne\hskip-\wd\tw@\mathord{\hskip\wd\tw@\hskip-\wd\@ne%
      {\vphantom{#3}}#1{#3}#2}}}
\def\@@rs#1#2{\mathinner{%
    \setbox\@ne\hbox{$\displaystyle{\vphantom{#2}}#1{#2}\m@th$}%
    \hskip\wd\@ne\mathord{\hskip-\wd\@ne%
      {\vphantom{#2}}#1{#2}}}}

\newcommand*{\norm}[1]{\mathinner{\Vert#1\Vert}}

\definecolor{notecolor}{cmyk}{0,1,1,.2}
\newcommand*\AM@notesname{Notes}

\makeatother

\modulolinenumbers[5]

\graphicspath{{02_figures/}}

\journal{The Journal of Adhesion}

\begin{document}

\begin{frontmatter}

\title{Investigation of the Peeling and Pull-off Behavior of Adhesive Elastic Fibers via a Novel Computational Beam Interaction Model}

\author{Maximilian J.~Grill\corref{cor1}}
\ead{grill@lnm.mw.tum.de}
\author{Christoph Meier}
\author{Wolfgang A.~Wall}

\address{Technical University of Munich, Institute for Computational Mechanics, Boltzmannstr.~15, 85748 Garching b.~M\"unchen, Germany}

\cortext[cor1]{Corresponding author}

\begin{abstract}%
This article studies the fundamental problem of separating two adhesive elastic fibers based on numerical simulation employing a recently developed finite element model for molecular interactions between curved slender fibers.
Specifically, it covers the two-sided peeling and pull-off process starting from fibers contacting along its entire length to fully separated fibers including all intermediate configurations and the well-known physical instability of snapping into contact and snapping free.
We analyze the resulting force-displacement curve showing a rich and highly nonlinear system behavior arising from the interplay of adhesion, mechanical contact interaction and structural resistance against (axial, shear and bending) deformation.
While similar to one-sided peeling studies from the literature, a distinct initiation and peeling phase can be observed, the two-sided peeling setup considered in the present work reveals the extended final pull-off stage as third characteristic phase.
Moreover, the influence of different material and interaction parameters such as Young's modulus as well as type (electrostatic or van der Waals) and strength of adhesion is critically studied.
Most importantly, it is found that the maximum force occurs in the pull-off phase for electrostatic attraction, but in the initiation phase for van der Waals adhesion.
In addition to the physical system behavior, the most important numerical aspects required to simulate this challenging computational problem in a robust and accurate manner are discussed.
Thus, besides the insights gained into the considered two-fiber system, this study provides a proof of concept facilitating the application of the employed model to larger and increasingly complex systems of slender fibers.
\end{abstract}


\end{frontmatter}

\section{Introduction}

Biopolymer fibers such as actin, collagen, keratin, cellulose and DNA, but also synthetic polymer, carbon and glass fibers or carbon nanotubes are ubiquitous examples for slender, deformable structures to be found on the scale of nano- to micrometers.
On these length scales, molecular interactions such as electrostatic or van der Waals (vdW) forces are of high relevance for the formation and functionality of the complex fibrous systems they constitute and in many of them adhesion plays a crucial role~\cite{French2010,Persson2003,israel2011,parsegian2005}.
To foster the understanding of such systems, which in turn allows for innovations in several fields from medical treatment to novel synthetic materials, there is an urgent need for powerful simulation tools.
This field of research has thus gained increasing attention over the last years and a review of the computational models for adhesion can be found in~\cite{Sauer2016}.
Concerning the above mentioned deformable fibers, it is desirable - and actually essential for practically relevant system sizes - to exploit the characteristic slenderness and describe the problem in a dimensionally reduced manner as a 1D Cosserat continuum which is well-known from beam theory.
In their recent contribution~\cite{GrillSSIP}, the authors of the present article have proposed the first computational model for molecular interactions, e.\,g.~vdW, repulsive steric, or electrostatic forces, between arbitrarily curved and oriented slender fibers undergoing large 3D deformations.
Its key idea is to formulate effective section-to-section interaction potential (SSIP) laws for the resultant interaction between a pair of fiber cross-sections, thus reducing the complexity of numerical integration to evaluate the total interaction potential from 6D to 2D.
This only enables predictive numerical simulations that are efficient yet accurate enough to cover practically relevant systems.

The objective of the present work is two-fold:
On the one hand, it employs the previously developed model to study the fundamental problem of separating two adhesive elastic fibers and to foster the understanding of the underlying physical mechanisms.
Specifically, it covers the peeling and pull-off process starting from fibers contacting along its entire length to fully separated fibers (and the reverse order) including all intermediate configurations and the well-known physical instability of snapping into contact and snapping free.
On the other hand, the present work serves as a proof of concept, facilitating future applications of the employed model to larger and increasingly complex systems of slender fibers.

This fundamental problem of separating two adhesive, elastic fibers seems to be intractable for purely analytical approaches due to the interplay of adhesion, repulsion and structural resistance against deformation, i.\,e., elasticity within the highly nonlinear regime of large deformations.
For the same reasons, it is challenging also for a computational approach and in fact the numerical treatment brings a number of further challenges with it (see~\cite{Sauer2011b} for a summary) which will be addressed in this work.
Several contributions to related problems can be found in the literature.
The adhesion of a Gecko spatula to solid surfaces has inspired the development of computational models to study and optimize the peeling behavior of thin elastic films and strips~\cite{Sauer2008,Sauer2011c,Sauer2014,Mergel2014,Mergel2014a}.
The corresponding model system of a beam interacting with a rigid half-space via the Lennard-Jones (LJ) potential shows certain similarities with the system of two deformable adhesive fibers to be studied in this article and we will return to this comparison in the discussion of our results.
Also the contribution~\cite{Sauer2013a}, where the peeling of two flexible strips is modeled via a 2D solid finite element formulation that combines a so-called cohesive zone model with a penalty contact formulation, will serve as source for comparison.
The problem studied in~\cite{Sauer2013a} differs from the one in the present work in terms of the specific geometry, boundary conditions, type of interactions and not at least in terms of the specific modeling and discretization approach.
It will thus be interesting to see how these differences carry over to the resulting force responses.
Two approaches to investigate the undesirable effect of clumping in fibrillar arrays used for bio-inspired dry adhesives are presented in~\cite{Ahmadi2014}.
The analytical approach aims to predict the critical fiber length leading to tip-tip contact by calculating the vdW force between assumed spherical tips of the fibers and applying it as a tip load in Euler-Bernoulli beam theory for a 2D cantilever.
This is complemented by a finite element approach using 2D solid elements and an effective inter-element vdW force based on the inverse-sixth power law.
The computational model proposed and applied in~\cite{Negi2018} focuses on the effect of inter-fiber adhesion at the level of 2D fiber networks.
It thus abstains from resolving the exact kinematics of adhesion and contact at the fiber scale and applies an effective energy gain per unit length of contacting parallel fiber segments and solves for the corresponding bundle segment lengths as additional unknowns.
Finally, an example for the electrostatic interaction of a double-clamped microbeam with a flat rigid electrode in a microelectromechanical system is given in~\cite{Shavezipur2011}.

The present study extends these previously published results for related problems with respect to the following aspects.
To the best of the authors' knowledge, this is the first study of the peeling behavior of two elastic fibers interacting via attractive electrostatic or vdW forces.
Compared to previous peeling studies, it also includes the pull-off phase, the intermediate regime around the physical instability of snapping into contact and snapping free, as well as the separated state of fibers.
Moreover, we investigate the scenario of peeling from both ends of the fibers, which turns out to show a similar behavior in the peeling phase yet a fundamentally different behavior in the pull-off phase as compared to the peeling from one side considered in the related problems of strip-rigid surface LJ interaction~\cite{Sauer2011c} and double strip debonding~\cite{Sauer2013a}.
Eventually, we analyze the resulting force-displacement curve revealing a rich, highly nonlinear system behavior and investigate the underlying physical mechanisms arising from the interplay of adhesion, mechanical contact interaction and structural resistance against (axial, shear and bending) deformation.
Furthermore, the influence of different material and interaction parameters such as Young's modulus as well as type (e.\,g.~electrostatic or van der Waals) and strength of adhesion on the resulting force-displacement relationship is studied by varying these parameters over two orders of magnitude.

In addition to the physical system behavior, the decisive numerical aspects required to simulate this challenging computational problem in a robust and accurate manner will be discussed in this article.
From a computational point of view, the major challenges resulting from the physical system characteristics are the delicate task of determining equilibrium configurations in the direct vicinity of the mentioned physical instability, the control of spatial discretization and numerical integration error such that the high gradients of short-range interaction potentials are represented with sufficient accuracy as well as the regularization of inverse power laws to eliminate the singularity at zero separation.
For the employed numerical model, such a regularization procedure has been proposed in~\cite{GrillSSIP}, which will turn out to enable a considerable increase in numerical robustness and efficiency at identical accuracy when applied to the highly challenging example of LJ interaction as considered in the present application.

The remainder of this article is structured as follows.
\secref{sec::methods} briefly recapitulates the employed physical models and numerical solution schemes as originally proposed in~\cite{GrillSSIP}.
In \secref{sec::num_ex_elstat_attraction_twoparallelbeams_peeling_pulloff}, the simulation of the peeling and pull-off process of two elastic fibers in case of electrostatic attraction will be presented and the underlying physical effects will be discussed in full detail.
\secref{sec::num_ex_vdW_attraction_twoparallelbeams_pulloff_from_contact} extends the numerical peeling experiment to the case of vdW adhesion and investigates the differences and similarities.
We conclude the article in~\secref{sec::conclusion_outlook} and give an outlook to promising aspects of potential future studies.

\section{Computational models and methods}\label{sec::methods}
The fundamental and unifying concept of all the methods employed in this work is the consistent dimensional model reduction from a 3D (Boltzmann) continuum to a 1D (Cosserat) continuum description applied to the slender fibers.
The kinematics of the latter type of model is uniquely defined via the positions and orientations of the fiber cross-sections, which is widely known from geometrically nonlinear beam theories.
Here, a set of well-established numerical formulations for the modeling of slender beams is combined with a novel computational model for (adhesive) molecular interactions between slender deformable fibers proposed in the authors' recent article~\cite{GrillSSIP}.
In this section, the individual methodological components will be presented in a concise manner and selected characteristics of special importance for this work will be highlighted.
The general solution strategy follows the one commonly used in nonlinear finite element frameworks for structural dynamics and roughly speaking consists of the following steps.
According to the principle of virtual work the weak form of the mechanical balance equations is derived and subsequently discretized in space and time.
Given a proper set of initial and boundary conditions, a load/time stepping scheme is applied and in every step the solution of the resulting discrete system of nonlinear equations is found iteratively by means of Newton's method.
The software package used for all the simulations in this work is the parallel, multi-physics, in-house research code BACI~\cite{BACI2018}.
Refer to~\cite{GrillSSIP} for details on the applied algorithms and numerical methods which go beyond the scope of the following overview.

\subsection{Elasticity of slender fibers}\label{sec::beam_theory}
The so-called geometrically exact 3D beam theory according to Reissner~\cite{reissner1981}, Simo~\cite{simo1985}, and Simo and Vu-Quoc~\cite{simo1986} is applied to describe the elastic deformation of the fibers.
It is known to be an accurate and efficient model including all the six deformation modes of axial tension, (2x) shear, torsion and (2x) bending.
In particular, it is applicable even for large deformations, finite 3D rotations, general cross-section shapes as well as arbitrary centerline shapes as stress-free reference configurations.
Here, the resulting system of nonlinear partial differential equations is discretized in space using beam finite elements according to Crisfield and Jeleni\'c~\cite{crisfield1999,jelenic1999}.
However, in contrast to the original work, we apply cubic Hermite polynomials for the spatial interpolation of the centerline.
The resulting $C^1$-continuity, i.\,e., smoothness at the element boundaries turns out to be of crucial importance in the context of numerical methods for beam-beam interactions such as contact~\cite{Meier2017b} or vdW adhesion~\cite{GrillSSIP}.
Refer to~\cite{Meier2017b} for the details and validation of this specific beam finite element formulation to be applied throughout this work.
At this point, we would like to point out that the numerical methods for beam-beam interactions to be presented in the following are independent of the specific beam formulation and have been applied to both Simo-Reissner and Kirchhoff-Love type formulations.
Note that the latter are known to be advantageous in the regime of high slenderness ratios where the underlying assumption of negligible shear deformation is met~\cite{Meier2017b,Meier2017c}.

\subsection{Inter-fiber adhesion: Section-to-section interaction potential approach}\label{sec::SSIP}
Inter-fiber adhesion is modeled by the so-called section-to-section interaction potential (SSIP) approach, which has been proposed in the authors' recent contribution~\cite{GrillSSIP}.
Starting from first principles for interactions of point charges or molecules, it has been derived specifically for interactions between curved slender fibers undergoing large deformations in 3D.
Following the idea of dimensional reduction from beam theory, it describes the effective interaction of the undeformable cross-sections by resultant SSIP laws~$\tilde{\tilde{\pi}}$ and thus reduces the numerical evaluation of the two-fiber interaction free energy~$\Pi_\text{ia}$ from two nested 3D integrals over both fibers' volumes~$V_i$ to two nested 1D integrals along the fibers' length dimensions~$l_i$:
\begin{align}
  \Pi_\text{ia} &= \iint_{V_1,V_2} \rho_1(\vx_1) \rho_2(\vx_2) \Phi(r) \dd V_2 \dd V_1\\
  &= \iint_{l_1,l_2} \; \underbrace{\iint_{A_1,A_2} \rho_1(\vx_1) \rho_2(\vx_2) \Phi(r) \dd A_2 \dd A_1}_{=: \; \tilde{\tilde{\pi}}(\vr_{1-2},\vpsi_{1-2})} \; \dd s_2 \dd s_1.
\end{align}
Here,~$\rho_i$ denotes the particle (volume) density of beam~$i \in \{1,2\}$ and~$\Phi(r)$ denotes the point-pair interaction potential as a function of the separation~$r=\norm{\vx_1-\vx_2}$.
In general, the SSIP law~$\tilde{\tilde{\pi}}$ will be a closed-form analytic function of the separation~$\vr_{1-2}$ of the centroid positions~$\vr_i$ and the relative rotation~$\vpsi_{1-2}$ between the cross-sections.
The specific expressions for the SSIP laws used for the studies in this work will be presented in the following Sections~\ref{sec::SSIP_elstat} and~\ref{sec::SSIP_vdW}.
In terms of the strategies to obtain a closed-form analytic expression for the SSIP law~$\tilde{\tilde{\pi}}$, analytical integration of the point pair potential is applied in~\cite{GrillSSIP}.
However, fitting postulated interaction laws to data from either experiments or all-atom molecular dynamics simulations of sample configurations is considered a promising alternative.

\subsubsection{Electrostatic interaction}\label{sec::SSIP_elstat}
The following SSIP law aims to describe the electrostatic interaction of nonconducting, circular cross-sections with constant surface charge densities~$\sigma_i$.
It is based on the monopole expression, i.\,e., the first term of the multipole expansion of each cross-section's surface charge distribution and reads
\begin{equation}\label{eq::SSIP_elstat}
  \tilde{\tilde{\pi}}_\text{elstat} = 2\pi R_1 \sigma_1 \, 2\pi R_2 \sigma_2 \, k_\text{elstat} \, d^{-1}.
\end{equation}
Here,~$d = \norm{ \vr_1 - \vr_2 }$ denotes the scalar centroid separation,~$R_i$ denote the cross-section radii, and~$k_\text{elstat}=(4\pi\epsilon_0\epsilon)^{-1}$ denotes the constant prefactor of Coulomb's law.
Note that this simple SSIP law only considers the net charge of the cross-sections as if it was concentrated at the centroid location and neglects any effect from rotations of the cross-sections, which would be included in the higher order terms of the multipole expansions.
However, at the scale of fiber-fiber interactions, this simple SSIP law proves to be surprisingly accurate as shown in the quantitative analysis in~\cite{GrillSSIP}, such that the inclusion of higher order terms seems not to be worth both the additional computational cost to evaluate them and the theoretical complexity associated with the rotational degrees of freedom required in this case.
Still, it is a viable option to enhance the accuracy of this SSIP law if it should be necessary at some point in the future.
Refer to~\cite{GrillSSIP} for the resulting virtual work contribution, spatial discretization and consistent linearization of this SSIP expression.

\subsubsection{Van der Waals interaction}\label{sec::SSIP_vdW}
Motivated by Hamaker-Lifshitz hybrid forms, point-pairwise summation - or integration - of the contributions from the fundamental inverse sixth power law is applied here.
Once again neglecting cross-section rotations, the simple SSIP law follows from the analytical 4D integration over two disks in parallel orientation, i.\,e., with parallel normal vectors as derived in~\cite{langbein1972}:
\begin{align}\label{eq::SSIP_vdW}
  \tilde{\tilde{\pi}}_\text{vdW} = \frac{ 3 \pi^2 }{ 256 } \,  \rho_1 \rho_2 \, \sqrt{ \frac{ 2 R_1 R_2 }{ R_1 + R_2 } } \, k_\text{vdW}\, g^{ -\frac{5}{2}}
\end{align}
Here,~$g = \norm{ \vr_1 - \vr_2 } - R_1 - R_2$ denotes the smallest surface separation also known as gap and~$k_\text{vdW}$ denotes the constant prefactor of the point-point interaction law, which can alternatively be replaced by Hamaker's constant~$A_\text{Ham} = \pi^2 \rho_1 \rho_2 k_\text{vdW}$ commonly used in this context.
Note that the accuracy of this simple SSIP law has been analyzed by means of analytical reference solutions for limiting cases on the scale of fiber-fiber interactions in~\cite{GrillSSIP}.
It shows that in contrast to the long-ranged electrostatic interaction considered in the preceding section, the two-fiber interaction potential is overestimated for small separations and concludes that the short range of vdW interactions requires to include the orientation of the cross-sections in order to obtain the correct power law exponent and thus enhanced accuracy.
As will be shown in~\secref{sec::results_num_ex_vdW_attraction_twoparallelbeams_pulloff_from_contact}, the model can however be calibrated by proper scaling of the parameter~$k_\text{vdW}$ to fit a given reference solution for a small range of separations~$g$, e.\,g.~around the equilibrium distance of the LJ potential, which is decisive for the global peeling force-displacement curve on the system level.
Throughout this work, the simple SSIP law for vdW adhesion stated above is thus considered a qualitative model, which is able to predict the correct qualitative behavior yet does not allow for reliable predictive quantitative analyses.
The influence of this known limitation on the results will be critically discussed in~\secref{sec::results_num_ex_vdW_attraction_twoparallelbeams_pulloff_from_contact} and the validity of the conclusions drawn will be secured.
Once again, refer to~\cite{GrillSSIP} for the resulting virtual work contribution, spatial discretization and consistent linearization of the presented SSIP expression.

\subsection{Steric repulsion of fibers}
Modeling the steric repulsion that prevents a penetration of distinct fibers has a long history in the field of computational contact mechanics and has first been addressed in~\cite{wriggers1997}.
The paradigm of these macroscopic continuum models is that the smallest surface separation or gap must be equal to or greater than zero which is formulated as an inequality constraint.
With the development of the novel SSIP approach to molecular interactions of fibers, an alternative modeling strategy has arisen, which is motivated by the rather microscopic perspective of LJ interactions between all material points in the slender continua.
In this work, both approaches will be applied, which asks for a brief assessment and comparison of the modeling approaches.

\subsubsection{Penalty-based model for beam contact}\label{sec::penalty_contact}
The beam contact formulation~\cite{meier2016} enforces the inequality constraint by means of a penalization of penetration, i.\,e., negative gap values.
In principle, arbitrary contact force-penetration laws may be applied, because the choice of a large enough penalty parameter value in any case ensures sufficiently small penetration values.
Here, a linear contact force law is enhanced by a quadratic regularization to obtain a smooth transition to zero contact forces for small positive gaps.
For each integration point along a fiber, the point-to-curve projection determines the closest point on the opposing fiber.
The resulting penalty potential (without the contribution from quadratic regularization) reads
\begin{equation}
  \Pi_\text{pen} = \frac{1}{2} \varepsilon \int_0^{l_1} \langle g \rangle^2 \dd s_1.
\end{equation}
Refer to~\cite{meier2016} for the corresponding expression for the virtual work, the steps of spatial discretization and consistent linearization as well as further algorithmic details.
Note that this line contact formulation can be combined with the point contact formulation~\cite{wriggers1997} in order to exploit the simplicity of point contact scenarios for the sake of efficiency as proposed in~\cite{Meier2017a}.

\subsubsection{Repulsive part of Lennard-Jones interaction}\label{sec::SSIP_repLJ}
The repulsive part of the LJ law deviates from the attractive vdW part only by the different (negative) exponent~$12$ instead of $6$.
It can thus be handled in the same way and the corresponding SSIP law reads
\begin{equation}
  \tilde{\tilde{\pi}}_\text{repLJ} =  \rho_1 \rho_2 \, \sqrt{ \frac{ 2 R_1 R_2 }{ R_1 + R_2 } } \, \tilde{k}_\text{repLJ}\, g^{ -\frac{17}{2}},
\end{equation}
where the constant prefactor~$\tilde{k}_\text{repLJ} \approx 5.30 \cdot 10^{-3} \, k_\text{repLJ}$ has been introduced in addition to the prefactor of the point-point interaction law~$k_\text{repLJ}$.
See~\secref{sec::SSIP} for a brief summary of the general SSIP approach and~\cite{GrillSSIP} for further details including the resulting virtual work contribution, spatial discretization and consistent linearization of this SSIP law.

\subsubsection{Brief assessment and comparison of the methods}
On the one hand, penalty-based models for beam contact are well-established formulations with optimized efficiency as well as robustness.
On the other hand, the SSIP approach is based on first principles in form of the LJ law and is thus expected to better resolve the actual contact force distributions, especially in the case of nano- to micro-scale applications.
This becomes obvious if we recall the purely heuristic nature of the penalty force law and the resulting (small) negative gap values, i.\,e., tolerated penetrations.
It will most likely depend on the specific application whether the associated model error has a significant or rather negligible influence on the results.
In order to answer this question with respect to the studies of this work, it seems most important to look at the adhesive force laws of~\secref{sec::SSIP} to be applied in combination with the models for steric repulsion.
The SSIP law~\eqref{eq::SSIP_elstat} modeling long-ranged electrostatic attraction is an inverse power law in the inter-axis separation~$d$ rather than the smallest surface separation~$g=d-R_1-R_2$ and thus expected not to be very sensitive to small changes in the gap values in case of contacting fibers~$g\approx0$.
On the contrary, the SSIP law~\eqref{eq::SSIP_vdW} is an inverse power law in the gap itself and therefore highly sensitive with respect to the gap~$g$.
Indeed, the thorough validation of both adhesion models using the example of two straight slender fibers in~\cite{GrillSSIP} as well as an unsuccessful attempt to use penalty beam contact in combination with vdW adhesion for the peeling simulation%
\footnote{The resulting peeling force values showed a noticeable unphysical dependence on both the type of the penalty force law and the value of the penalty parameter~$\varepsilon$.
}
to be presented in~\secref{sec::num_ex_vdW_attraction_twoparallelbeams_pulloff_from_contact} confirm these considerations.
Moreover, refer to~\cite{GrillSSIP} for a detailed discussion of the importance to correctly resolve the regime of small gap values by means of a suitable regularization strategy to remedy the inherent singularity of the vdW SSIP law~\eqref{eq::SSIP_vdW} for zero separation~$g \to 0$.

To conclude, the choice of a proper computational model for steric repulsion between contacting fibers is closely linked to the type of adhesion and most probably even depends on the specific application.
For the reasons outlined above, the penalty-based line contact formulation will be applied together with the SSIP law for electrostatic attraction throughout~\secref{sec::num_ex_elstat_attraction_twoparallelbeams_peeling_pulloff} whereas the SSIP approach based on the repulsive part of the LJ law will be applied in combination with the SSIP law for vdW adhesion in~\secref{sec::num_ex_vdW_attraction_twoparallelbeams_pulloff_from_contact}.

\section{Electrostatic attraction}\label{sec::num_ex_elstat_attraction_twoparallelbeams_peeling_pulloff}
The system considered in the simulation consists of two initially straight and parallel, deformable fibers that attract each other due to their constant surface charge of opposite sign.
Its setup is kept as simple as possible to allow for an isolated and clear analysis of the physical effects as well as the characteristics of the computational model.
Note that the static equilibrium configurations as a result of different surface charge densities at a fixed, large separation of the fibers have already been studied in the authors' recent contribution~\cite{GrillSSIP}.
The study of the peeling and pull-off process to be presented in this section can thus be considered the more advanced continuation of this analysis in order to obtain the full picture of the system behavior from fibers clinging together along their entire length to fully separated fibers.

\subsection{Setup and parameters}\label{sec::num_ex_elstat_attraction_twoparallelbeams_peeling_pulloff_setup}
\begin{figure}[htpb]%
  \centering
  \subfigure[Problem setup]{
    \def\svgwidth{0.13\textwidth}
    \input{num_ex_elstat_attraction_twoparallelbeams_pulloff_from_contact_problem_setup.pdf_tex}
    \hspace{0.02\textwidth}
    \label{fig::num_ex_elstat_attraction_twoparallelbeams_pulloff_from_contact_problem_setup}
  }
  \subfigure[Quasi-static force-displacement curve. Force values to be interpreted as multiple of a reference point load that causes a deflection of~$l/4$ if applied at the fiber midpoint.]{
    \includegraphics[width=0.8\textwidth]{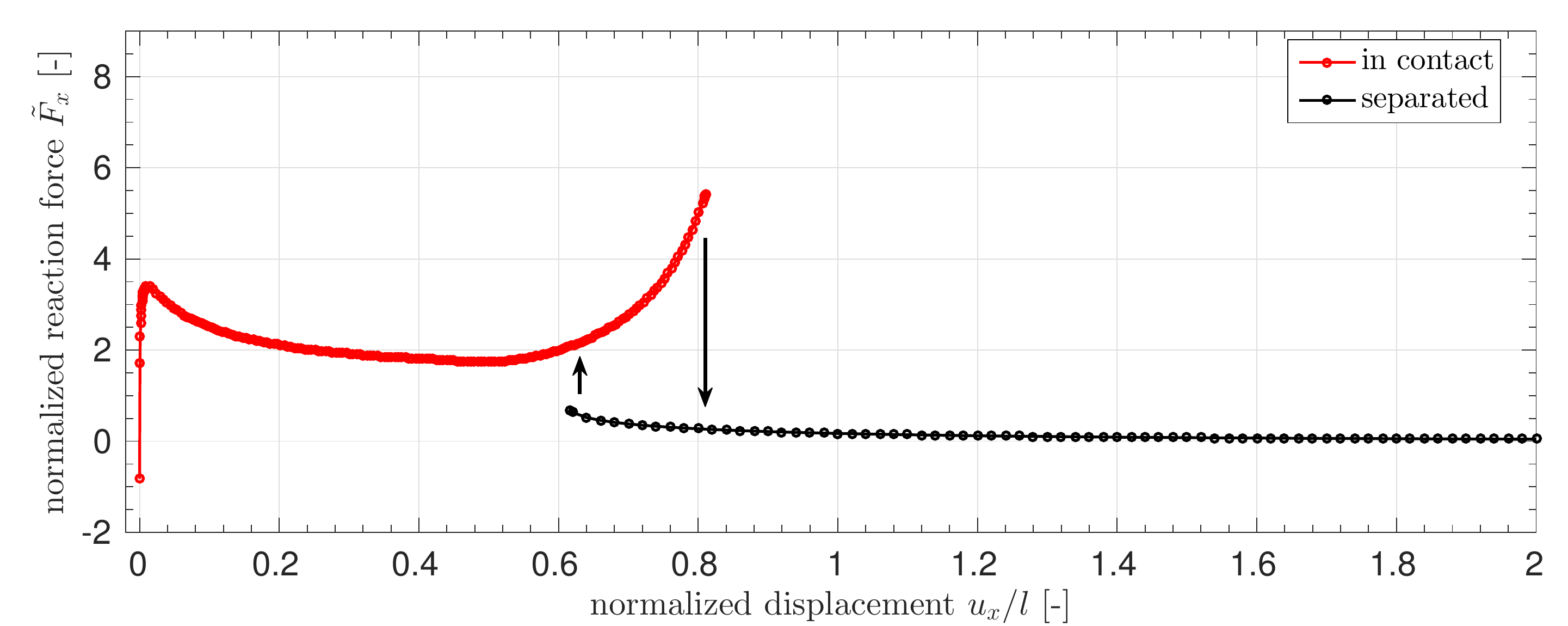}
    \label{fig::num_ex_elstat_attraction_twoparallelbeams_pulloff_force_over_displacement}
  }
  \caption{Numerical peeling and pull-off experiment with two adhesive elastic fibers.}
  \label{fig::num_ex_elstat_attraction_twoparallelbeams_pulloff}
\end{figure}

As shown in~\figref{fig::num_ex_elstat_attraction_twoparallelbeams_pulloff_from_contact_problem_setup}, two straight fibers of length~$l=5$ are aligned with the global~$y$-axis and are simply supported at their endpoints.
Additionally, both fibers are restricted to move only within the~$xy$-plane and rotate only around the global~$z$-axis.
The fibers have a circular cross-section with radius~$R=0.02$, which results in a slenderness ratio of~$\zeta = l/R = 250$.
The initial configuration is chosen such that the fiber surfaces are in contact, i.\,e., the fiber centerlines are placed with an inter-axis separation~$d_0=2R$.
Starting from this initial state, a horizontal displacement~$u_x$ will be prescribed to both supports of the right fiber and the total resulting horizontal force~$F_x = F_x^{tr} + F_x^{br}$ will be analyzed.
Cross-section area, area moments of inertia and shear correction factor are computed using standard formula for a circle.
A hyperelastic material law with Young's modulus~$E=10^5$ and Poisson's ratio~$\nu=0.3$ is applied.

Electrostatic interaction is accounted for via the SSIP approach as presented in~\secref{sec::SSIP_elstat}.
Both fibers are nonconducting with a constant surface charge density of~$\sigma_1=1.0$ and~$\sigma_2=-1.0$, respectively.
The strength of attraction is controlled via the prefactor~$k$ of the Coulomb law~$\Phi(r) = k r^{-1}$ and set to~$k=0.1$ for the first part of this study.
An analysis of the effect of parameter value variation will be content of~\secref{sec::varying_youngs_modulus}.

The repulsive contact forces are modeled by means of the frictionless line penalty contact formulation as presented in~\secref{sec::penalty_contact}.
In particular, a quadratically regularized linear penalty law with line penalty parameter~$\varepsilon = 100$ and regularization parameter~$\bar g = R/10$ is applied here.
Recall at this point, that the surface-to-surface separation - or gap - $g$ is defined as~$g=d-R_1-R_2$ and thus negative gap values indicate a penetration, whereas positive gap values imply physical separation of the bodies.
The applied regularization therefore means that contact forces smoothly increase in the regime of small positive gaps~\mbox{$0<g<\bar g$}, starting from a force value as well as slope of zero at~$g=\bar{g}$.
As it will turn out, these exemplarily chosen parameter values lead to a slightly positive equilibrium spacing~$g_{eq} \approx R/100$, where the repulsive contact forces balance the adhesive electrostatic forces.
The initial state of the system with zero separation~$g_0 = d_0-2R = 0$ along the entire length of the fibers will thus lie in the compressive, i.\,e., repulsive regime with negative reaction force values~$\tilde F_x<0$.

For the following quantitative analysis, we use a fine spatial discretization of~$n_\text{ele}=16$ Hermitian Simo-Reissner beam elements (see~\secref{sec::beam_theory} for details) per fiber to ensure that the discretization error has no perceptible effect on the results.
For the same reason, we choose a high number of Gauss points for the numerical integration of contact as well as electrostatic forces.
Two integration segments per element with ten Gauss points each are used for electrostatic forces and~$20$ integration segments per element with five Gauss points each are used for contact forces.
This turns out to be fine enough to not change the presented results perceptibly.
A closer look at mesh refinement and the undesirable effect of too coarse meshes follows at the end of this section.

Obviously, the competition of attractive and repulsive forces of which both are strongly deformation-dependent leads to a complex system behavior.
The associated nonlinearity and stiffness are highly demanding with respect to solving the nonlinear system of equations.
We thus apply Newton's method in combination with a displacement increment control as described in~\cite{GrillSSIP}.
The upper bound of the displacement increment per iteration is chosen as~$|\Delta u|_\text{max}= R/2$ here, which prevents an undetected crossing of fibers.

\subsection{Results}
\begin{figure}[htpb]%
  \centering
  \subfigure[$u_x/l=0$]{
    \includegraphics[width=0.4\textwidth]{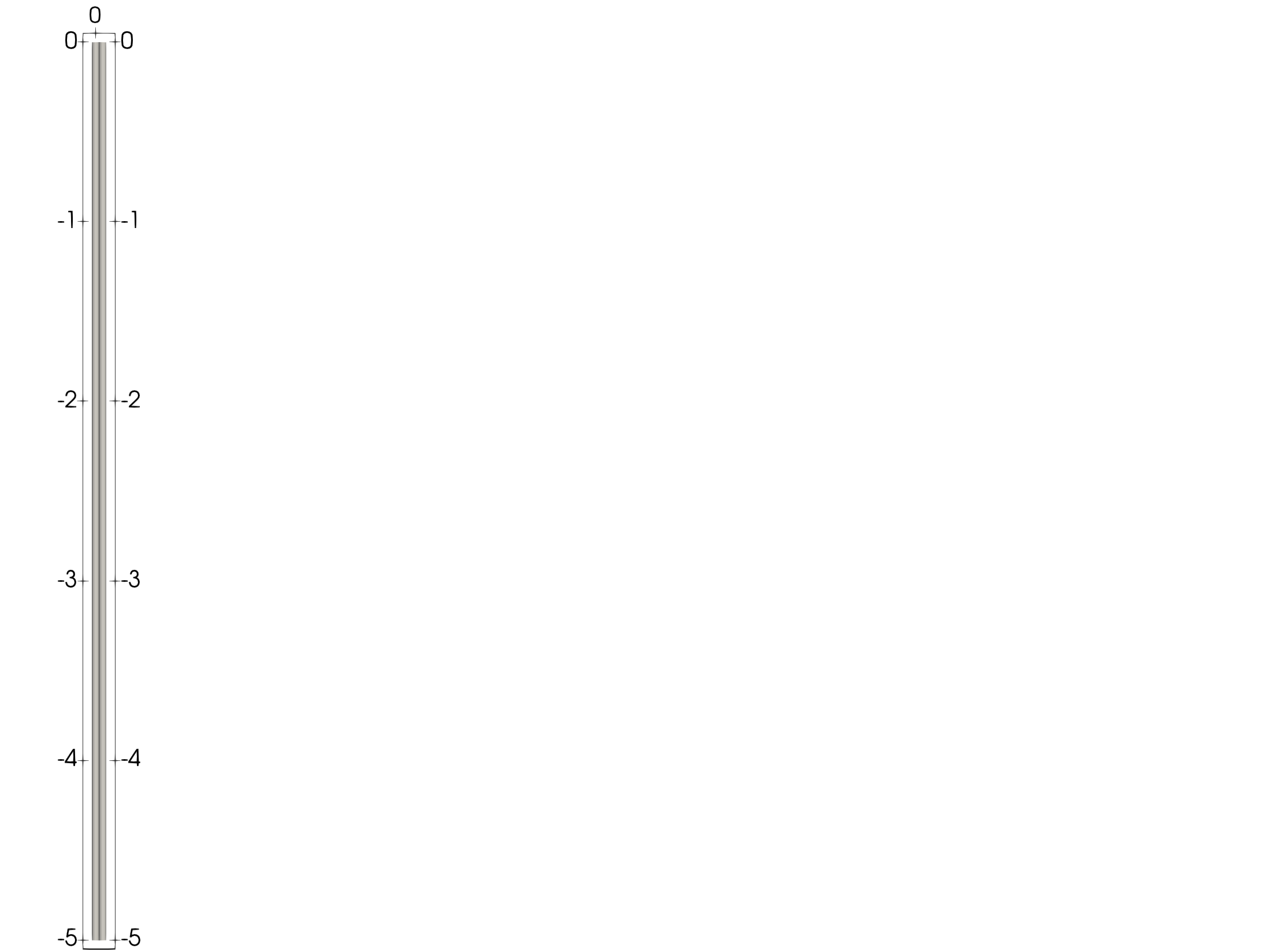}
    \label{fig::num_ex_elstat_attraction_twoparallelbeams_pulloff_initial_config}
    \hspace{-4.5cm}
  }
  \subfigure[$u_x/l=0.01$]{
    \vspace{3cm}
    \includegraphics[width=0.385\textwidth]{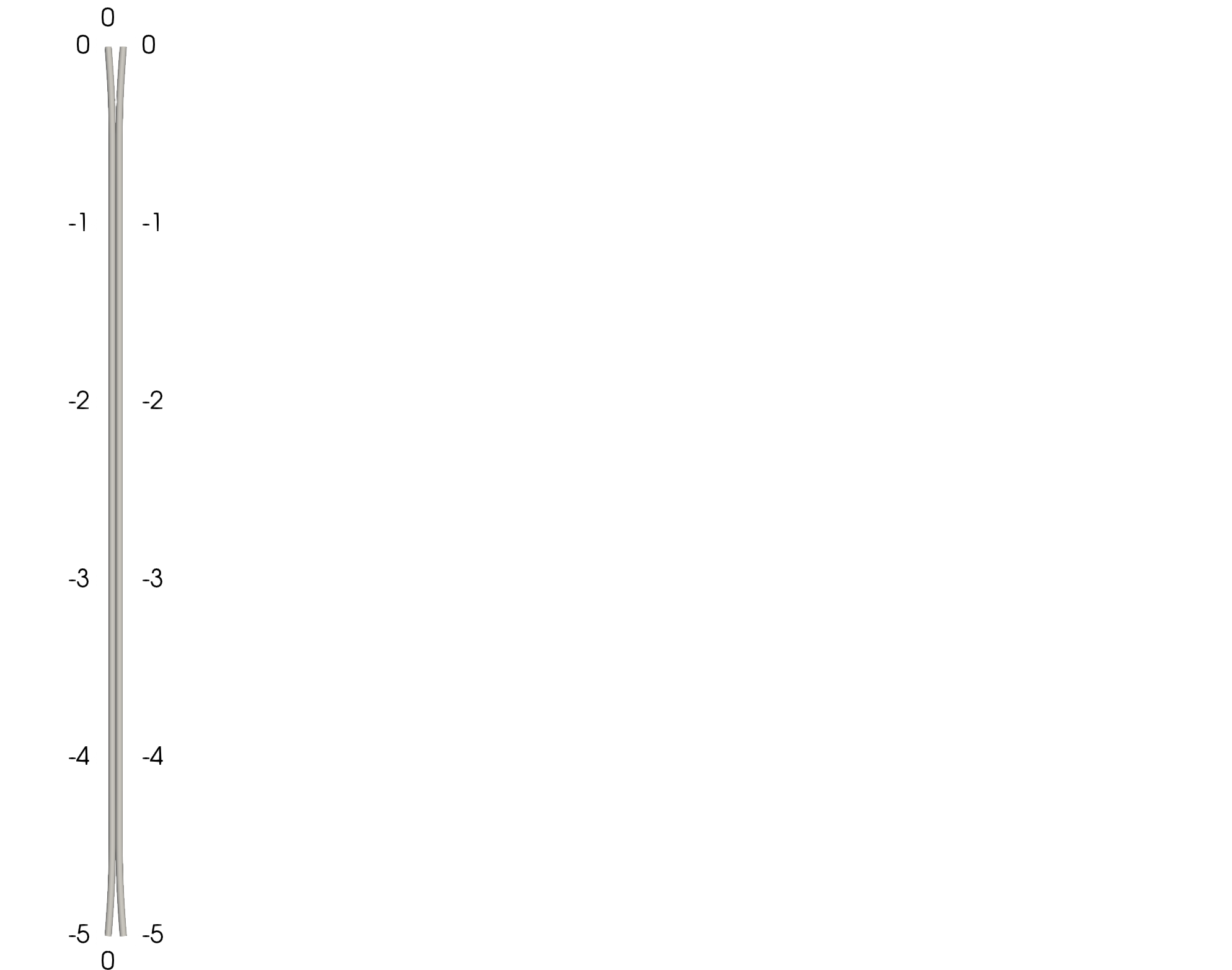}
    \label{fig::num_ex_elstat_attraction_twoparallelbeams_pulloff_separation0_045}
    \hspace{-4.5cm}
  }
  \subfigure[$u_x/l=0.2$]{
    \includegraphics[width=0.4\textwidth]{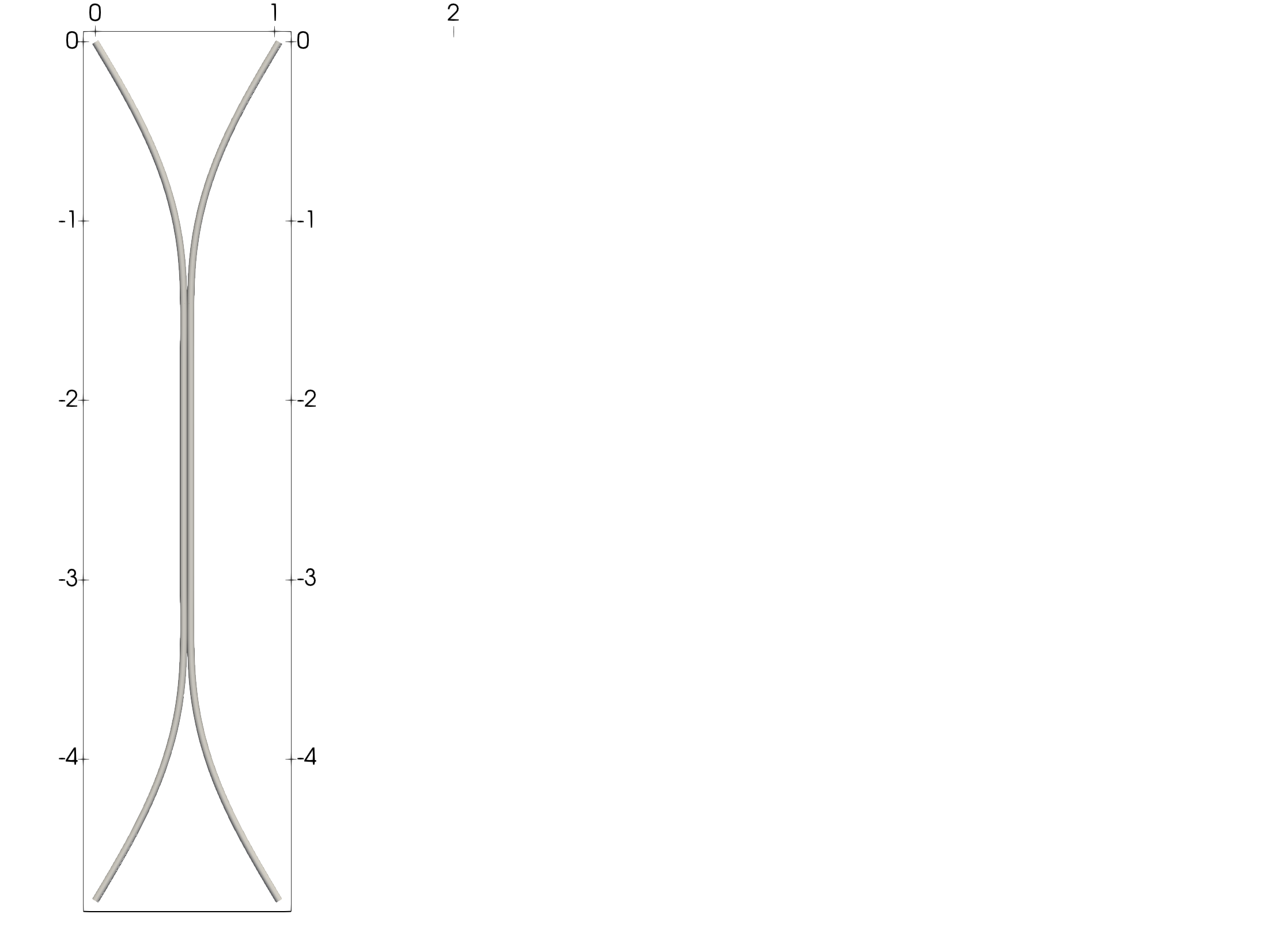}
    \label{fig::num_ex_elstat_attraction_twoparallelbeams_pulloff_separation1_00}
    \hspace{-4cm}
  }
  \subfigure[$u_x/l=0.4$]{
    \includegraphics[width=0.4\textwidth]{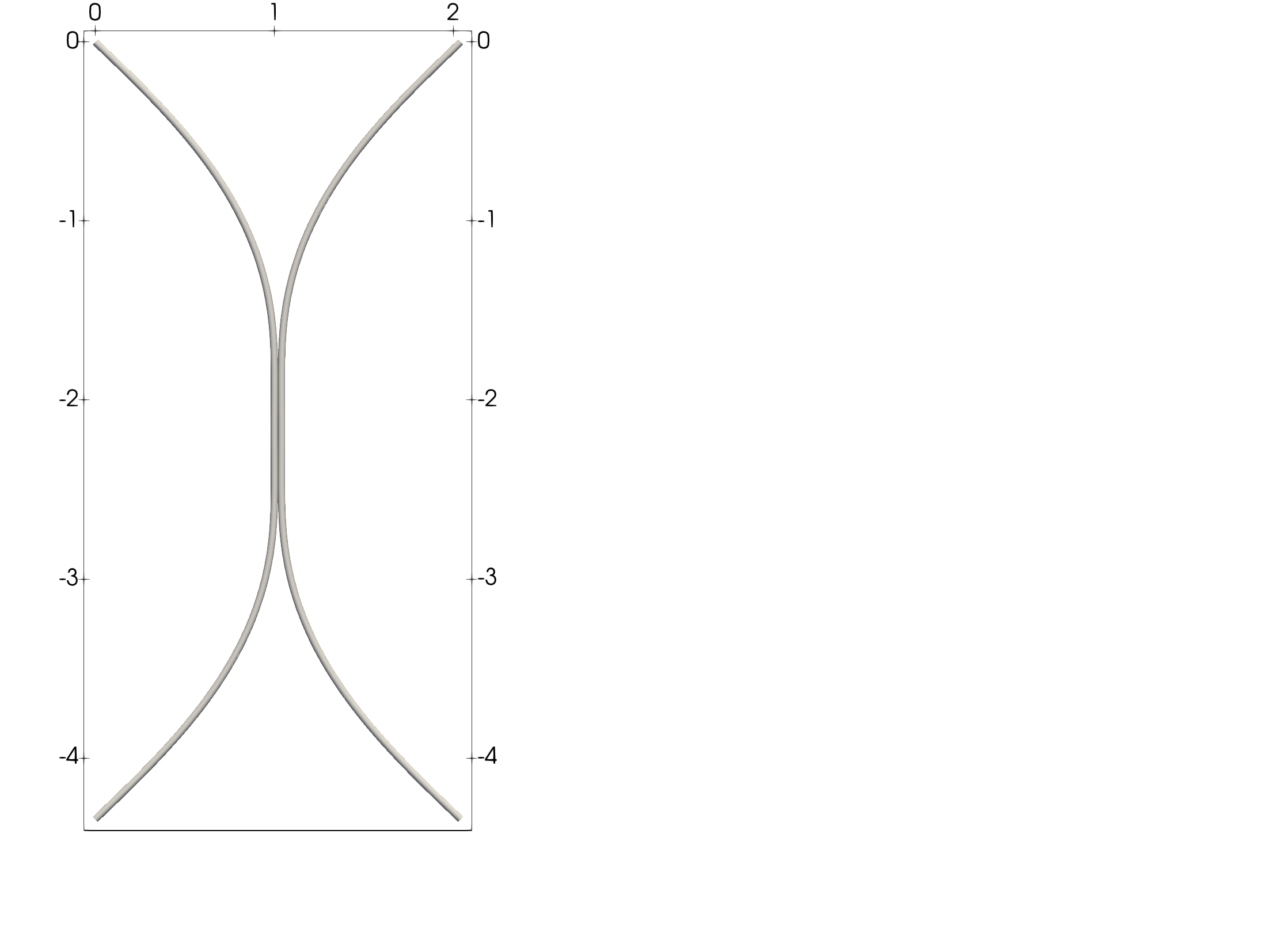}
    \label{fig::num_ex_elstat_attraction_twoparallelbeams_pulloff_separation2_00}
    \hspace{-4cm}
  }
  \subfigure[$u_x/l=0.5$]{
    \includegraphics[width=0.4\textwidth]{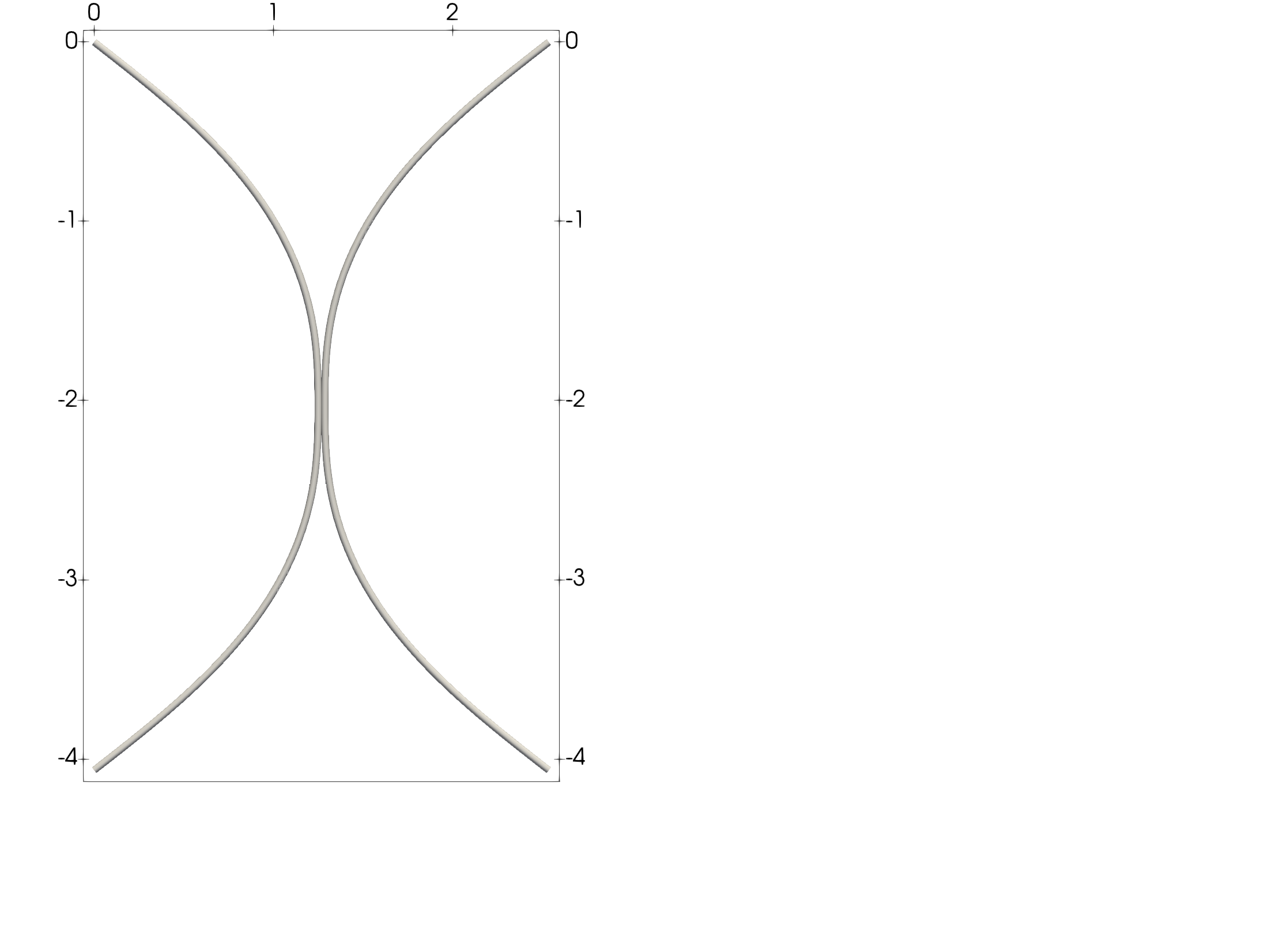}
    \label{fig::num_ex_elstat_attraction_twoparallelbeams_pulloff_separation2_50}
    \hspace{-3cm}
  }
  \subfigure[$u_x/l\approx0.62$: in contact]{
    \includegraphics[width=0.21\textwidth]{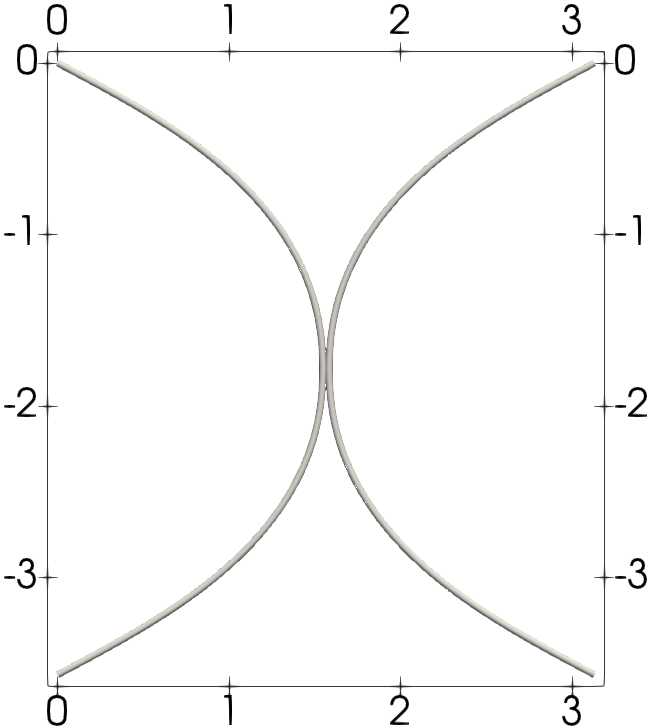}
    \label{fig::num_ex_elstat_attraction_twoparallelbeams_pulloff_separation3_089_incontact}
    \hspace{0.25cm}
  }
  \subfigure[$u_x/l=0.8$: in contact]{
    \includegraphics[width=0.26\textwidth]{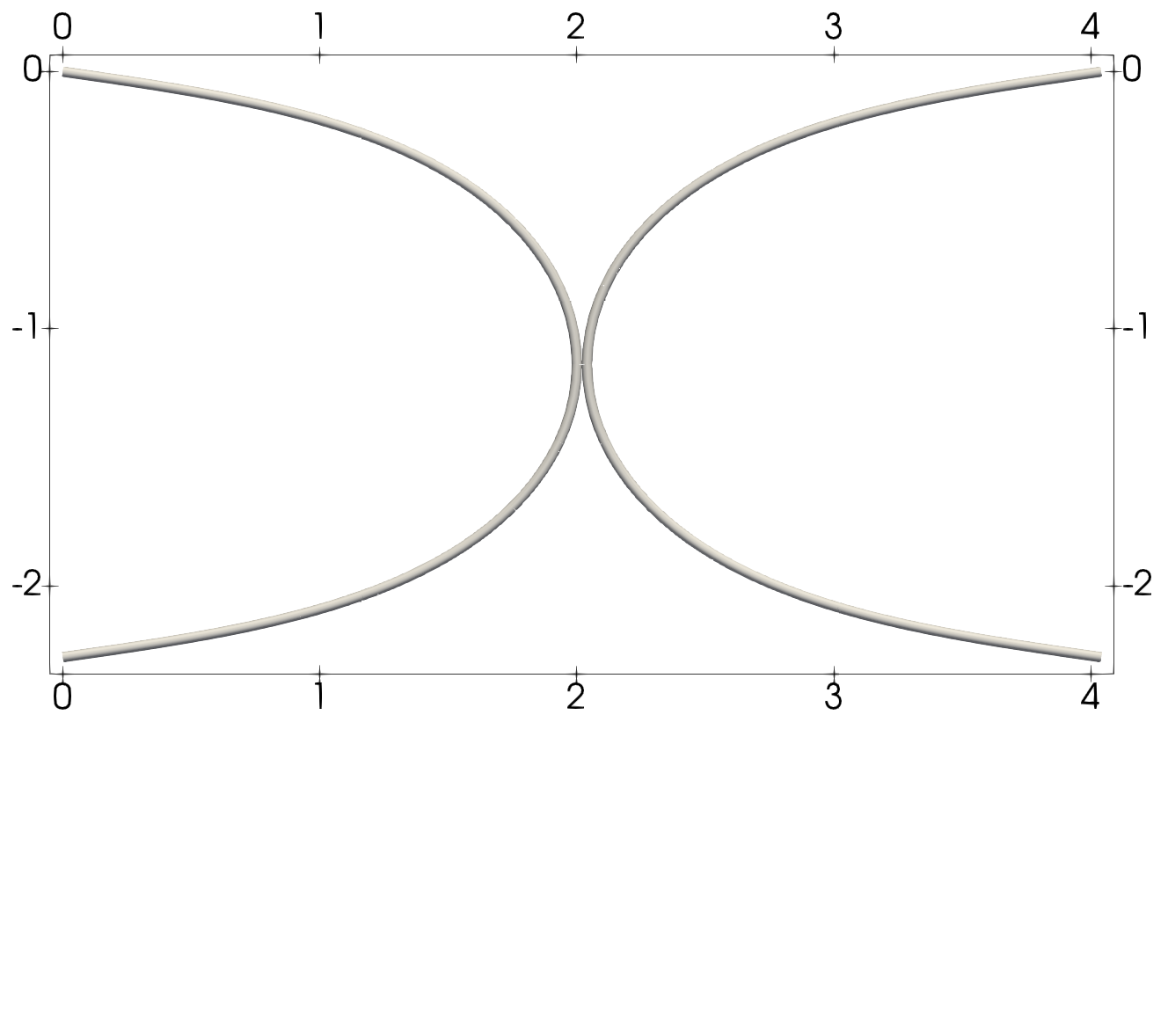}
    \label{fig::num_ex_elstat_attraction_twoparallelbeams_pulloff_separation4_00_incontact}
    \hspace{0.25cm}
  }
  \subfigure[$u_x/l\approx0.8105$: rightmost red data point in Fig.~\ref{fig::num_ex_elstat_attraction_twoparallelbeams_pulloff_force_over_displacement}, ultimately before fibers would snap free]{
    \includegraphics[width=0.26\textwidth]{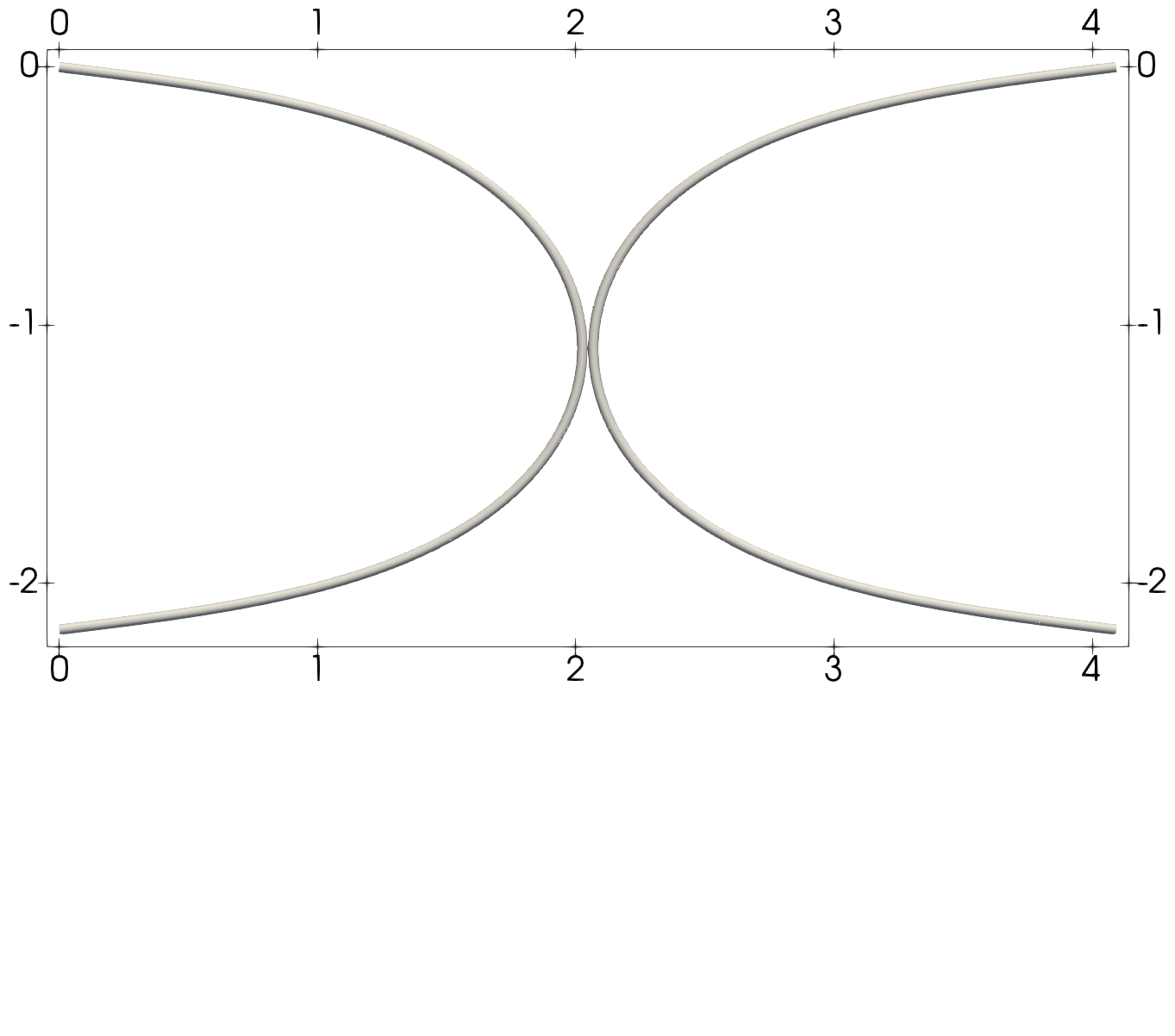}
    \label{fig::num_ex_elstat_attraction_twoparallelbeams_pulloff_finalstatebeforesnapfree}
    \hspace{2cm}
  }
  \subfigure[$u_x/l\approx0.62$: leftmost black data point in Fig. \ref{fig::num_ex_elstat_attraction_twoparallelbeams_pulloff_force_over_displacement}, ultimately before fibers would snap in contact]{
    \includegraphics[width=0.21\textwidth]{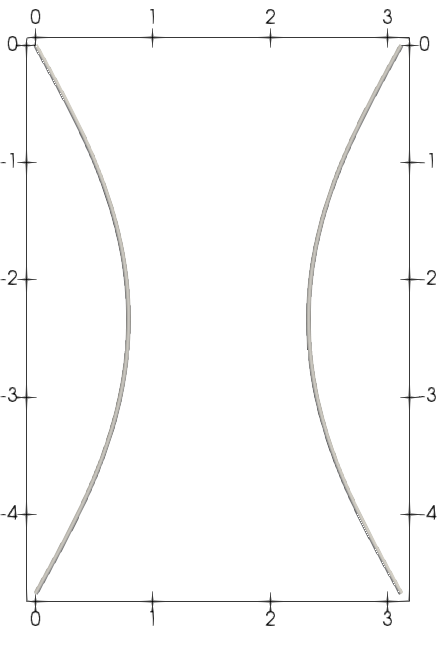}
    \label{fig::num_ex_elstat_attraction_twoparallelbeams_pulloff_separation3_08}
  }
  \subfigure[$u_x/l=0.8$: separated]{
    \includegraphics[width=0.4\textwidth]{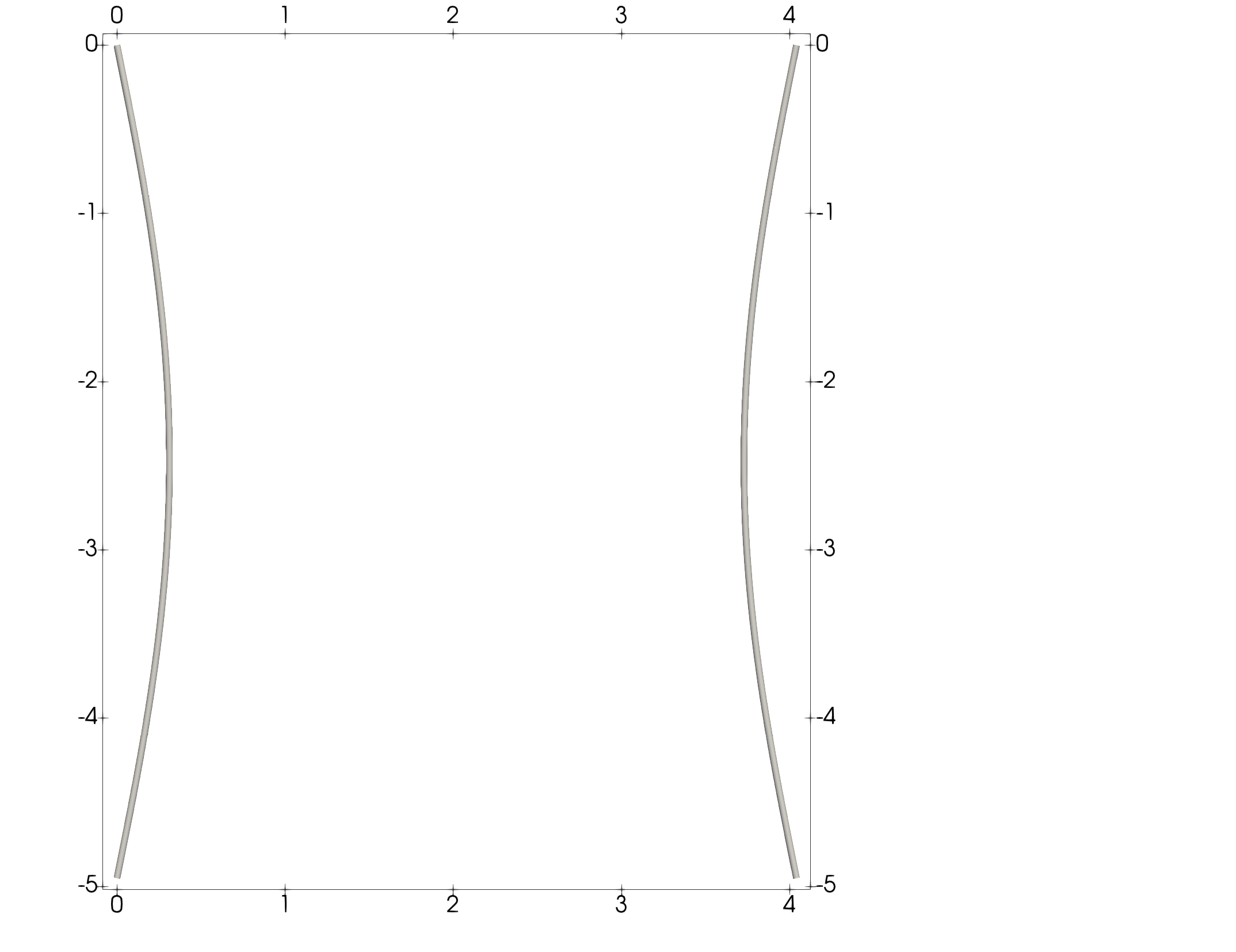}
    \label{fig::num_ex_elstat_attraction_twoparallelbeams_pulloff_separation4_00}
    \hspace{-2cm}
  }
  \subfigure[$u_x/l=1.2$: separated]{
    \includegraphics[width=0.4\textwidth]{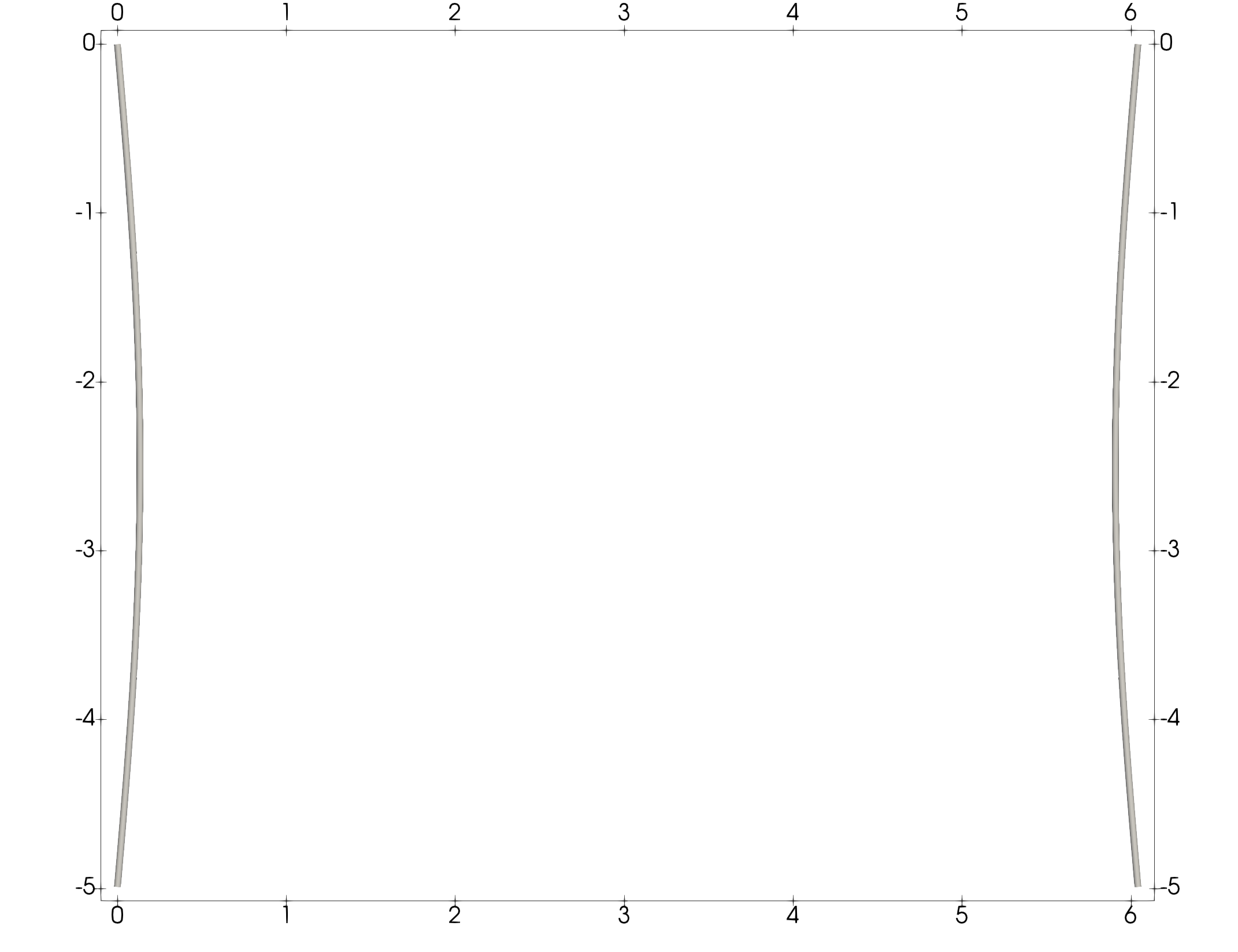}
    \label{fig::num_ex_elstat_attraction_twoparallelbeams_pulloff_separation6_00}
  }
  \caption{Simulation snapshots of the peeling and pull-off experiment of two adhesive elastic fibers. Note that all configurations are symmetric with respect to a vertical as well as horizontal axis. This also implies $F_x^\text{tl}=F_x^\text{bl}=-F_x^\text{tr}=-F_x^\text{br}$, $F_y\equiv0$. Note that (a) - (h) show configurations on the red branch, (i) - (k) on the black branch of~\figref{fig::num_ex_elstat_attraction_twoparallelbeams_pulloff_force_over_displacement}, respectively. }
  \label{fig::num_ex_elstat_attraction_twoparallelbeams_pulloff_snapshots}
\end{figure}
The resulting force-displacement curve shown in \figref{fig::num_ex_elstat_attraction_twoparallelbeams_pulloff_force_over_displacement} reveals a rich and interesting system behavior.
Most obvious, there are two distinct branches of static equilibrium configurations that do not merge.
On the one hand, starting from zero displacement and illustrated in red, we see the branch where the fibers are in contact and on the other hand, for large separations, there is another branch depicted in black where the fibers are separated.
The transition between both states as indicated with arrows will always be a dynamic process and will be discussed in further detail later on.
Simulation snapshots for some characteristic displacement values are provided in \figref{fig::num_ex_elstat_attraction_twoparallelbeams_pulloff_snapshots}.

Let us first look at the left part of the force-displacement plot where the fibers are in contact.
Note that the force values are normalized and to be interpreted as multiple of a reference point load that causes a deflection of l/4 if applied at the fiber midpoint.
At zero displacement and thus zero separation of the fiber surfaces, the fibers repel each other as a resistance against penetration which results in a negative reaction force~$\tilde F_x \approx -0.8$.
From the second data point onwards, we identify the tensile regime with positive, rapidly increasing force values.
At~$u_x/l \approx 0.01$, the force reaches a local maximum value of~$\tilde F_x \approx 3.9$ and then decreases until it reaches a local minimal value of~$\tilde F_x \approx 1.74$ at~$u_x/l \approx 0.5$.
Upon further displacement, the force increases again until the fibers suddenly snap free at some point.
The exact point of snapping free strongly depends on the dynamics of the system and can thus not be determined in this quasi-static analysis.
However, it is very interesting to see that for a certain range of separation~$0.6 \lessapprox u_x/l \lessapprox 0.8$ two different static equilibrium configurations exist - one with contacting fibers and one with separated fibers.
See also the corresponding simulation snapshots in~\figref{fig::num_ex_elstat_attraction_twoparallelbeams_pulloff_separation4_00_incontact} and~\figref{fig::num_ex_elstat_attraction_twoparallelbeams_pulloff_separation4_00}, respectively.
The largest horizontal separation of the supports for which we could find a static equilibrium configuration with the fibers being still in contact, i.\,e., the rightmost red data point yields~$\tilde F_x \approx 5.4$ at~$u_x/l \approx 0.81$.
Beyond this point, Newton's method fails to converge.
This is reasonable because the nearest solution of the nonlinear system of equations looks similar to~\figref{fig::num_ex_elstat_attraction_twoparallelbeams_pulloff_separation4_00} whereas the last converged configuration and thus initial guess for Newton's method is~\figref{fig::num_ex_elstat_attraction_twoparallelbeams_pulloff_finalstatebeforesnapfree}.

The second branch of static equilibrium with separated fibers is much more intuitive and some qualitative aspects have already been discussed in the scope of the authors' previous contribution~\cite{GrillSSIP}, where the resulting equilibrium configurations for varying attractive strength at a fixed, large separation~$u_x/l=1$ have been studied.
Here, we also present the quantitative analysis of the resulting force values as a function of the displacement and particularly include the intermediate range of separations~$0.6 \lessapprox u_x/l \lessapprox 0.8$.
Due to the long range of electrostatic forces, we still observe a perceptible force and deflection for separated fibers.
Nevertheless, as expected, force values decay and approach zero for large separation.
Most interesting however is once again the range of~$0.6 \lessapprox u_x/l \lessapprox 0.8$.
Here, no static equilibrium could be found with Newton's method as fibers tend to jump into contact and so the equilibrium states are unstable.
Instead, we conducted dynamic simulations with artificial viscous damping forces and waited until the system had reached its steady state%
\footnote{The steady state has been defined in a way that the magnitude of every velocity and acceleration component in the system has fallen below a threshold value of~$10^{-10}$.}.
The method used to compute the viscous forces has been proposed in \cite{Cyron2010} and models the interaction of a semi-flexible filament with a quiescent background fluid.
In this manner, we could determine further (unstable) static equilibrium configurations in the range of~$0.616 \lessapprox u_x/l \lessapprox 0.9$.
As discussed earlier for the pull-off, also the exact point of jump-into-contact will depend on the dynamics of the system.

\begin{figure}[htpb]%
  \centering
  \subfigure[$u_x/l=0$]{
    \includegraphics[width=0.5\textwidth]{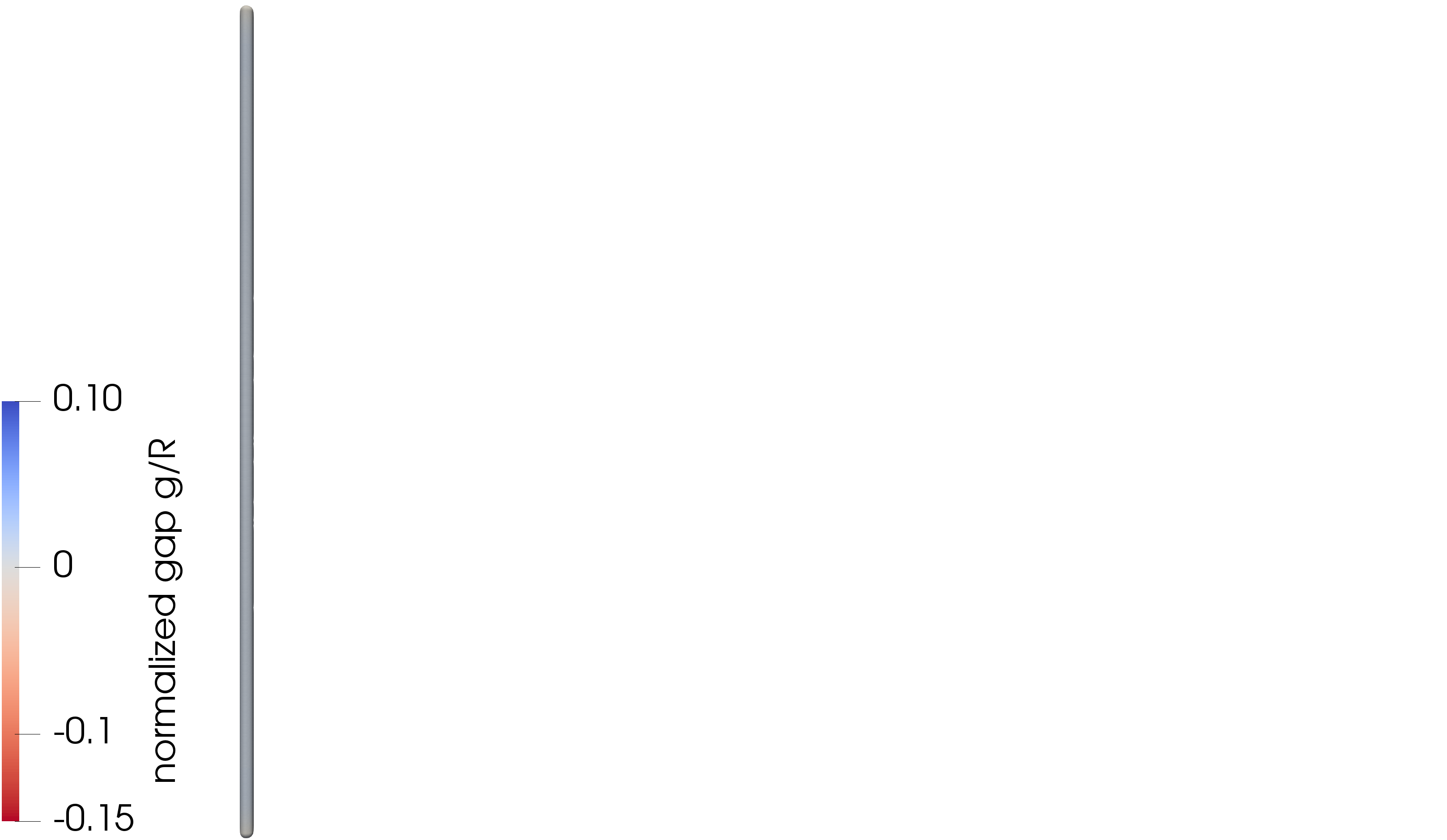}
    \label{fig::num_ex_elstat_attraction_twoparallelbeams_pulloff_gaps_initial_config}
    \hspace{-6cm}
  }
  \subfigure[$u_x/l=0.01$]{
    \hspace{-1cm}
    \includegraphics[width=0.5\textwidth]{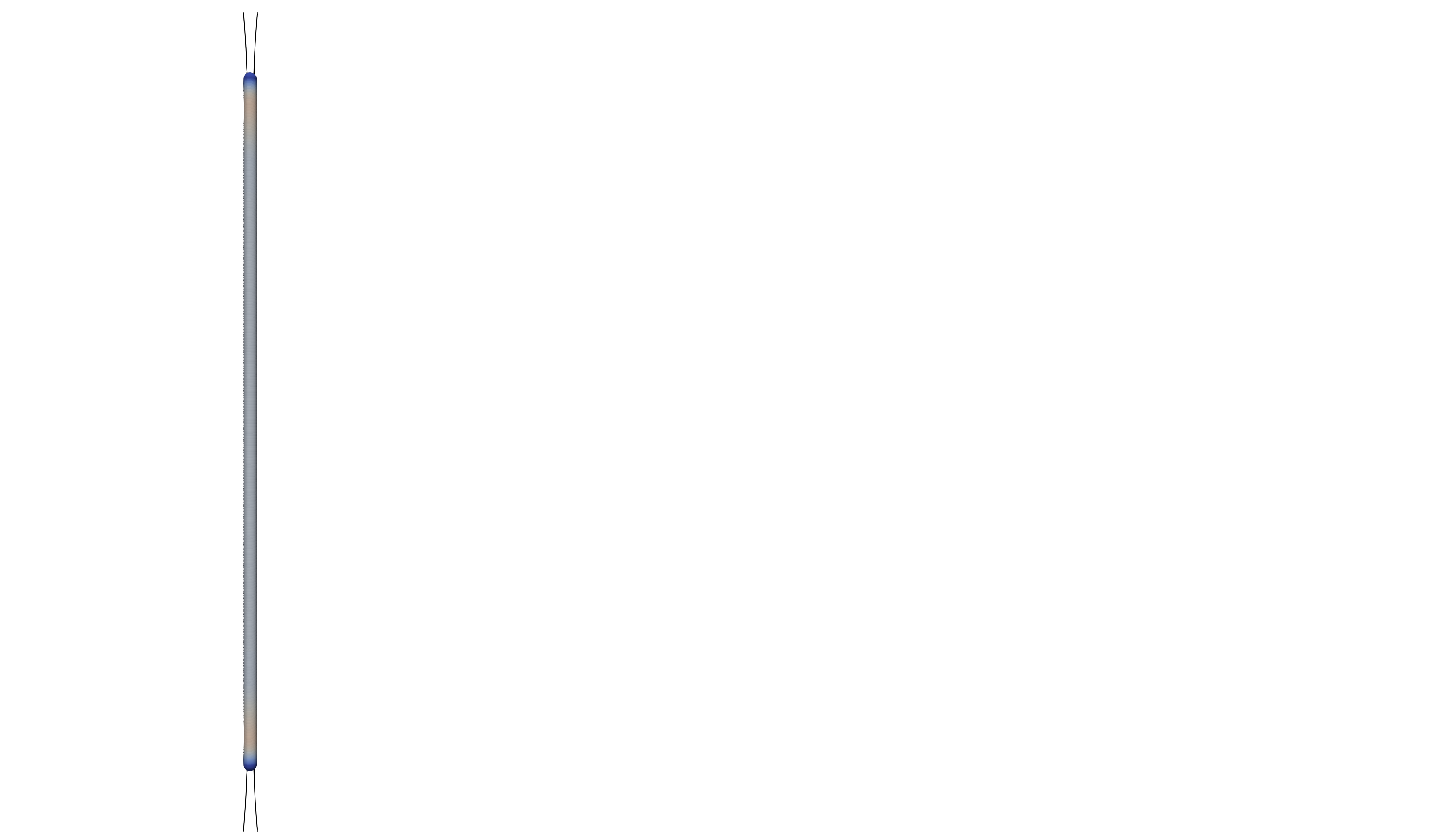}
    \label{fig::num_ex_elstat_attraction_twoparallelbeams_pulloff_gaps_separation0_045}
    \hspace{-5.5cm}
  }
  \subfigure[$u_x/l=0.2$]{
    \hspace{-1cm}
    \includegraphics[width=0.5\textwidth]{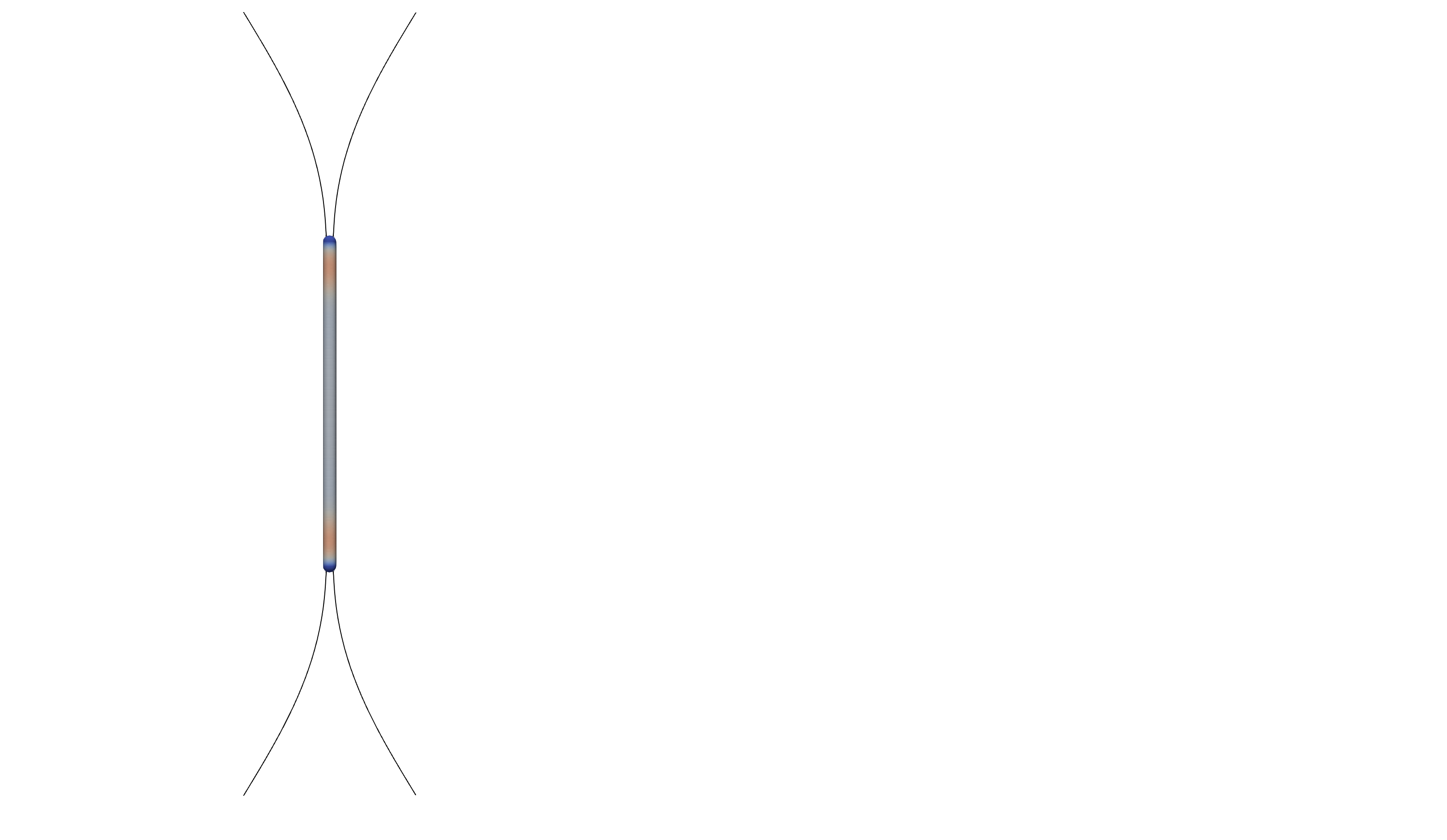}
    \label{fig::num_ex_elstat_attraction_twoparallelbeams_pulloff_gaps_separation1_01}
    \hspace{-5.5cm}
  }
  \subfigure[$u_x/l \approx 0.61$: maximal penetration]{
    \hspace{-1cm}
    \includegraphics[width=0.5\textwidth]{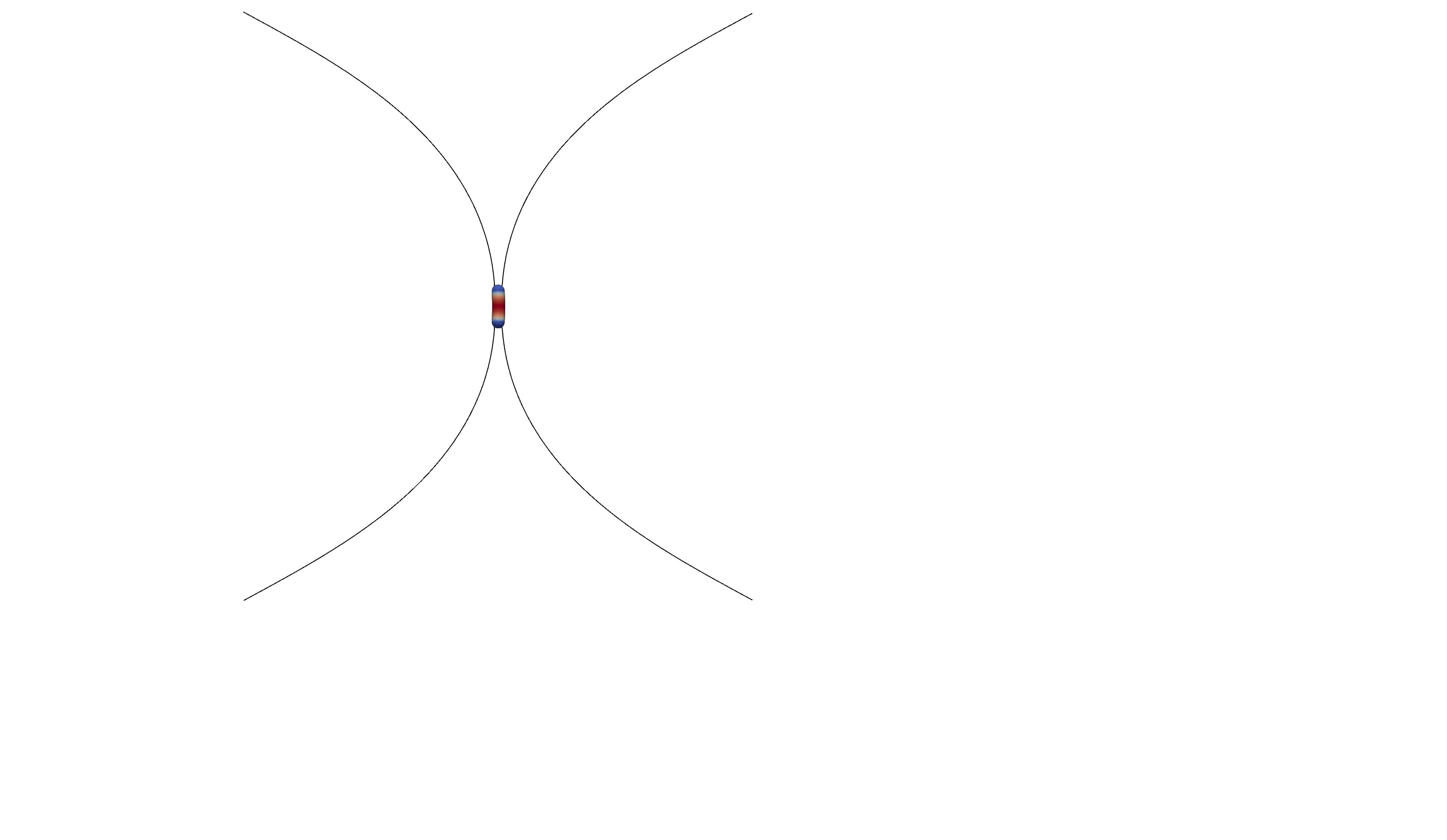}
    \label{fig::num_ex_elstat_attraction_twoparallelbeams_pulloff_gaps_separation3_06_maximal_penetration}
    \hspace{-4cm}
  }
  \subfigure[$u_x/l\approx0.8105$: ultimately before the fibers would snap free]{
    \hspace{-1cm}
    \includegraphics[width=0.5\textwidth]{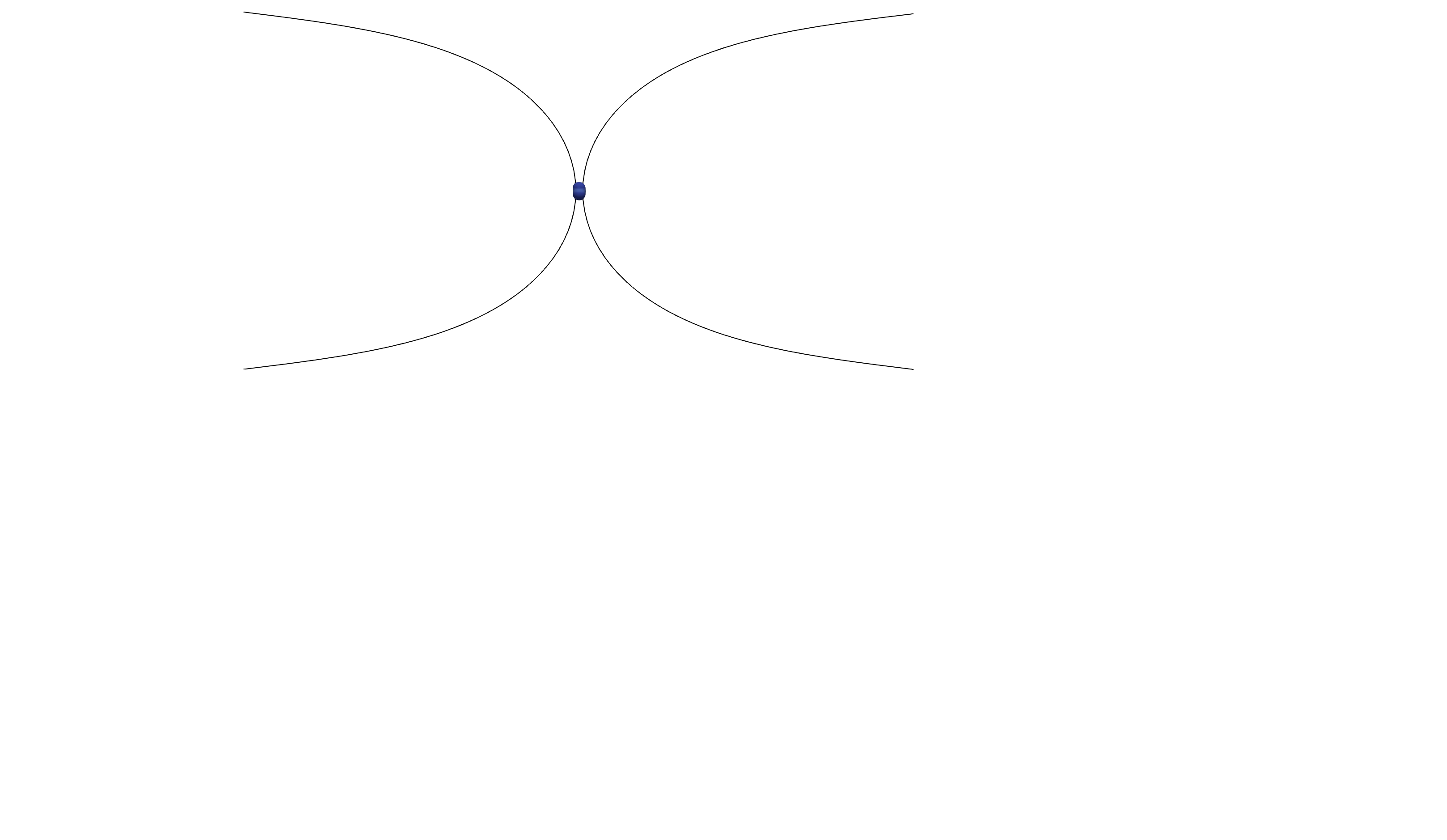}
    \label{fig::num_ex_elstat_attraction_twoparallelbeams_pulloff_gaps_finalstatebeforesnapfree}
    \hspace{-3cm}
  }
  \caption{Visualization of surface-to-surface separation during the peeling and pull-off experiment of two adhesive elastic fibers. Fibers represented by their centerlines and active contact pairs illustrated as spheres located half way between interacting points on fibers. Color indicates gap values normalized with the fiber radius~$g/R$. Negative gap values indicate penetration.}
  \label{fig::num_ex_elstat_attraction_twoparallelbeams_pulloff_gaps}
\end{figure}

Since the separation of the fiber axes/surfaces is a non-trivial result of the competing attractive and repulsive forces, it is worth to have a closer look at the gap values.
From \figref{fig::num_ex_elstat_attraction_twoparallelbeams_pulloff_snapshots}, one can already tell that contacting fibers neither show large penetrations nor a visible gap between the surfaces and as such meet our expectations.
In \figref{fig::num_ex_elstat_attraction_twoparallelbeams_pulloff_gaps}, points of active contact are visualized as spheres and coloring indicates the gap values~$g/R$, i.\,e., normalized with respect to the fiber radius.
We find that $g/R \gtrapprox -0.15$, i.\,e., the maximal penetration is less than~$15\%$ of the fiber radius throughout the entire simulation from zero displacement to separation.
In the context of penalty-based models for beam contact, this is considered a reasonable small value and may be interpreted as a model for the cross-section deformation, which is otherwise precluded in the applied beam theory.
Note also that the maximal penetration only occurs for a very short time interval at~$u_x/l \approx 0.61$ in the spatially confined region around the fiber midpoints.

\paragraph{Remark on the reference force value}
At~$u_x/l=0.5$, the deflection of the fiber midpoint(s) is the same as in the experiment to determine the reference force used for normalization. The corresponding force value of~$\tilde F_x\approx1.7$ can thus be interpreted as a surplus of deformation as compared to the simple reference experiment with a point load applied to the midpoint.

\subsection{Discussion}
In the following discussion, we will elaborate on a deeper understanding of the observed system behavior.
Particularly, the underlying mechanisms of the complex force-displacement curve from zero displacement to full separation of the fibers shall be elucidated.
To this end, one has to discuss the attractive as well as repulsive forces between both fibers along with the internal elastic forces and moments in the fibers.
Generally, the prevalent mode of fiber deformation is bending, i.\,e., curvature of fiber axes.
Axial elongation and shear deformation remain small and no torsional deformation occurs due to the planar setup.
Concerning the interaction forces, it is most plausible to look at a net interaction line load acting on each of the fibers.
To begin with, we basically identify three distinct phases that are discussed individually in the following.

\begin{description}
  \item [\normalfont{\textit{Phase a) initiation of fiber deformation and detachment}} $0 < u_x/l < 0.01$:]
  The displacement of the fiber endpoints initiates a curvature in a locally confined region towards the endpoints of the fiber.
  Accordingly, the first surface point pairs begin to separate, leave the equilibrium spacing and cause an adhesive force.
  Since the fiber axes are almost parallel and the fibers show a resistance against bending deformation, a considerable number of surface point pairs must detach at once.
  This explains the steep increase of the pulling force in the leftmost part of the force-displacement diagram.
  The major part in the middle of the fibers however remains parallel in an equilibrium state of balanced contact and adhesive forces.
  The two limiting cases would be rigid fibers, where the entire length of the fiber needs to detach at once and rope-like fibers, i.\,e., with negligible bending stiffness, where one point pair after another could detach.
  It is thus a competition of elastic deformation and electrostatic adhesion, that will be further analyzed in a parameter study with varying Young's modulus at the end of this section.
  \figref{fig::num_ex_elstat_attraction_twoparallelbeams_pulloff_separation0_045} illustrates the configuration corresponding to the local force maximum which can be considered as the end of this initial phase.
  \item [\normalfont{\textit{Phase b) peeling}} $0.01 < u_x/l < 0.5$:]
  Subsequently, the contact zone continuously decreases as more and more surface point pairs detach which can be identified as peeling.
  Especially in the beginning of this peeling phase, the opening angle between the fiber axes increases and likewise the pulling force decreases.
  This is due to the known effect, that a larger angle requires less point pairs to detach at the same time.
  Additionally, the lever arm in form of the already detached, free part of the fiber becomes longer, such that the reaction force at the supports decreases.
  The combination of the increasing opening angle and the decreasing effective stiffness of the longer free fiber parts are presumably the most important effect in this phase.
  From the perspective of fiber deformation, the radius of curvature increases as compared to the initially induced strong local curvature.
  Still, the middle part of the fibers remains parallel whereas the end parts are bent.
  Interestingly, a closer look at the gap values shown in \figref{fig::num_ex_elstat_attraction_twoparallelbeams_pulloff_gaps_separation1_01} reveals that the resulting centerline shape resembles a very slight ``w``.
  In the course of the peeling phase, the opening angle as well as the free fiber length and likewise the pulling force approaches a constant value.
  This constant peeling force over displacement is well-known from thin film peeling in the limit of zero bending stiffness~\cite{Sauer2011c} as predicted analytically by~\cite{Kendall1975}.
  \item [\normalfont{\textit{Phase c) pull-off}} $0.5 < u_x/l < 0.8$:]
  The end of the peeling and begin of the pull-off phase can be identified as the point from which on the pulling force increases again.
  As can be seen from~\figref{fig::num_ex_elstat_attraction_twoparallelbeams_pulloff_separation2_50} and \ref{fig::num_ex_elstat_attraction_twoparallelbeams_pulloff_gaps_separation3_06_maximal_penetration}, the centerline shape now resembles a ''c``, i.\,e., is convex.
  Accordingly, the contact zone diminished to a short region in the middle of the fibers that can not easily be peeled any further because it would require a strong local curvature of the fibers.
  During the entire pull-off phase, the adhesive forces acting on the fibers change only marginally because there is not much change in the mutual distance of the most important closest parts of the fibers.
  However, the repulsive contact forces decrease and therefore the net interaction force increases considerably.
  So the remarkable increase in the external pulling force before the fibers finally snap free results from the compensation of the diminishing contact forces.
  The maximum value of the tensile force that is required during the separation of adhesive bodies is commonly referred to as \textit{pull-off force} and is of highest relevance in many practical applications.
  Looking at this phase from the different perspective of elastic deformation, the fibers undergo a high curvature in the middle part and increasing axial tension towards the endpoints in order to conform with the ever increasing separation~$u_x/l$, such that in the end each fiber resembles a ''u``-shape%
  \footnote{This shape will be even more pronounced for a smaller value of Young's modulus, cf.~\figref{fig::num_ex_elstat_attraction_twoparallelbeams_static_pulloff_comparison_youngsmodulus_finalstatebeforesnapfree}.}.
  The high axial stiffness of the fibers leads to an increase in the reaction force until it ultimately reaches the maximum value that can be transferred by the adhesive connection.
\end{description}

When comparing these results to those obtained for the related scenario of the one-sided peeling of a thin elastic film adhering to a rigid surface via vdW forces in the previous study~\cite{Sauer2011c}, a number of similarities can be identified.
In both of the aforementioned phases a) and b), the resulting force-displacement curves have the same characteristic shape, i.\,e., a steep initial slope towards a sharp force peak followed by gradually decreasing force values eventually approaching a plateau-like regime of almost constant peeling forces thus representing a strongly nonlinear, deformation-dependent system behavior that arises from the interplay of elasticity, adhesion and mechanical contact interaction.
This also holds true for the double strip peeling modeled via 2D solid elements and a cohesive zone model studied in~\cite{Sauer2013a}, which underlines the universality of both the results itself as well as of the thorough analysis of the underlying physical mechanisms.
Neither the differences in the physical origin (and computational modeling) of adhesion nor the slightly different setup with respect to loading and support, i.\,e., the different boundary conditions, seem to change this fundamental system response during peeling and the initiation thereof.
We will get back to this topic in the discussion of the results for the parameter variation in~\secref{sec::varying_youngs_modulus} and for the case of vdW adhesion in~\secref{sec::results_num_ex_vdW_attraction_twoparallelbeams_pulloff_from_contact}.

Turning to the differences in the results as well as its reasons, the most obvious observation from~\figref{fig::num_ex_elstat_attraction_twoparallelbeams_pulloff_force_over_displacement} is the pronounced pull-off phase described above.
As compared to the mentioned previous studies, this can be attributed to the application of pulling forces at both ends of the fibers, which results in a two-sided instead of a one-sided peeling.
While the pull-off displacement and force in case of one-sided peeling strongly depend on the properties and modeling of the fiber endpoints, the two-sided peeling setup shifts the focus to the interaction of the fibers' middle parts.
The observed significant increase of the force over an extended range of displacement values $0.5 < u_x/l < 0.8$ before snapping free is therefore considered to be characteristic for the two-sided peeling setup, which is an important finding with implications concerning the assessment of real-world systems of adhesive fibers.

Following up on this, the most striking fact is that the global force maximum occurs at the end of this pull-off phase c), ultimately before snapping free.
As the results for the case of vdW attraction will show, this appears to be a distinguishing feature of the long range of electrostatic attraction considered here.
In view of the fact that the global force maximum will be the decisive feature in biological as well as bio-inspired synthetic dry adhesives, a clear understanding of how the maximum force value and the corresponding displacement or, more generally, system state depends on the elementary system properties is certainly of highest importance.
Thus, the following section investigates the effect of varying the principal parameters, and subsequently the fundamentally different type of adhesion arising from vdW interactions will be studied in~\secref{sec::num_ex_vdW_attraction_twoparallelbeams_pulloff_from_contact}.

\subsection{Influence of the strength of adhesion and Young's modulus}\label{sec::varying_youngs_modulus}
Having studied the interplay of elasticity and adhesion for one set of parameters, we now want to look at the effect of parameter variation.
As has been noted already by Sauer~\cite{Sauer2011} in the scenario of peeling a thin film with finite bending stiffness from a rigid surface, the decisive parameter is the ratio of Young's modulus and adhesive strength.
The simulations conducted in the scope of this section confirmed that this holds true also for the case of two adhesive elastic fibers studied here.
In the following, we thus leave the prefactor of the point potential law~$k=0.1$ unchanged and vary Young's modulus~$E$ by a factor of~$0.1$ and~$10$, respectively, thus covering a range of two orders of magnitude.

\begin{figure}[htpb]%
  \centering
  \subfigure[Quasi-static force-displacement curves. Force values to be interpreted as multiple of a reference point load that causes a deflection of~$l/4$ if applied at the fiber midpoint. Here, we use the same reference force value - the one which is obtained for Young's modulus~$E=10^{5}$ - for all the three curves.]{
    \includegraphics[width=0.4\textwidth]{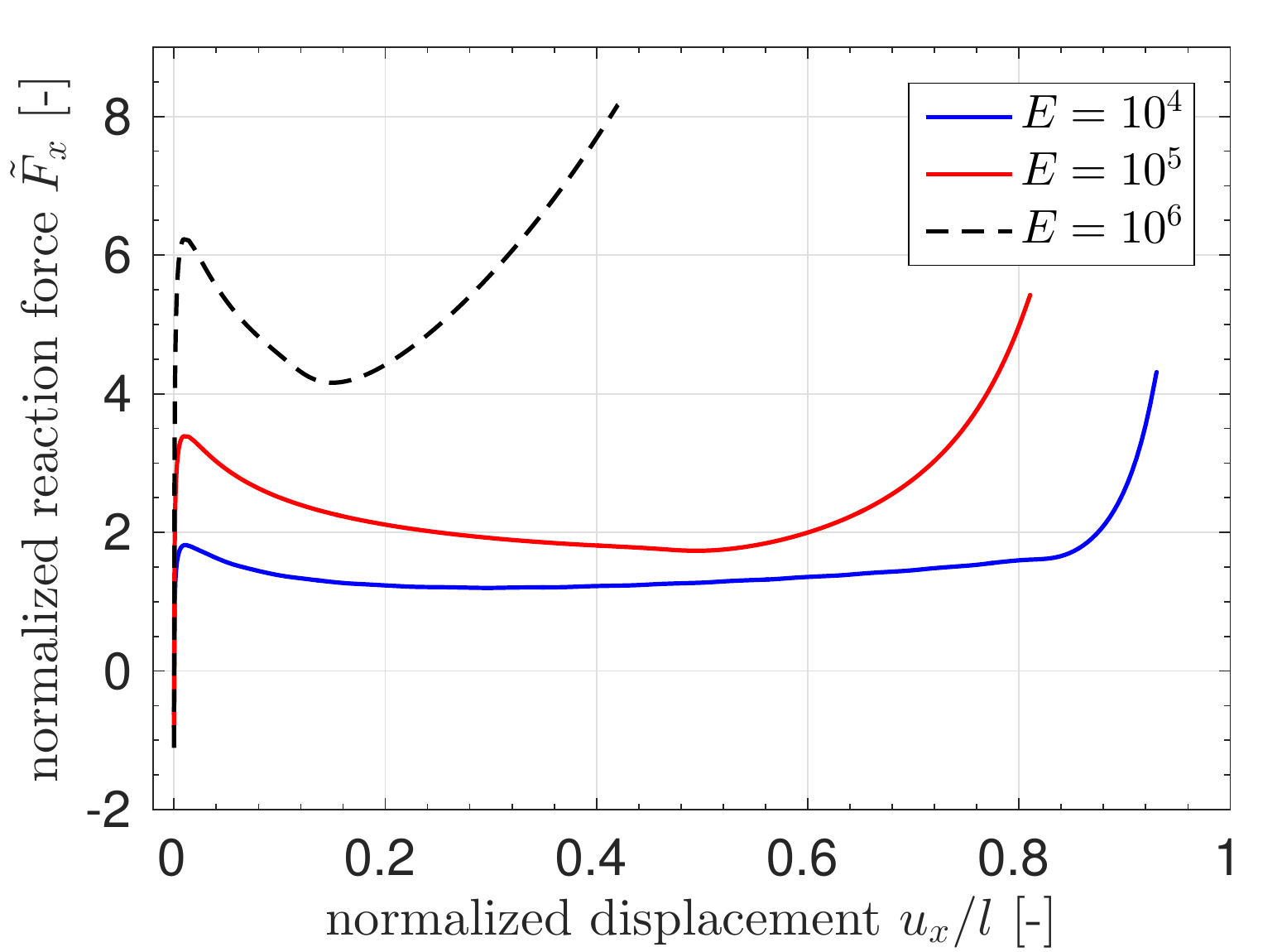}
    \label{fig::num_ex_elstat_attraction_twoparallelbeams_static_pulloff_reactionforce_over_displacement_comparison_youngsmodulus}
  }%
  \hspace{1cm}
  \subfigure[Quasi-static force-displacement curves with alternative normalization: Force values are again to be interpreted as multiple of a reference point load that causes a deflection of~$l/4$ if applied at the fiber midpoint, but here we use an individual reference force value for each curve, which is obtained for the corresponding value of Young's modulus \mbox{$E=10^4, 10^5, 10^6$}, respectively.]{
    \includegraphics[width=0.4\textwidth]{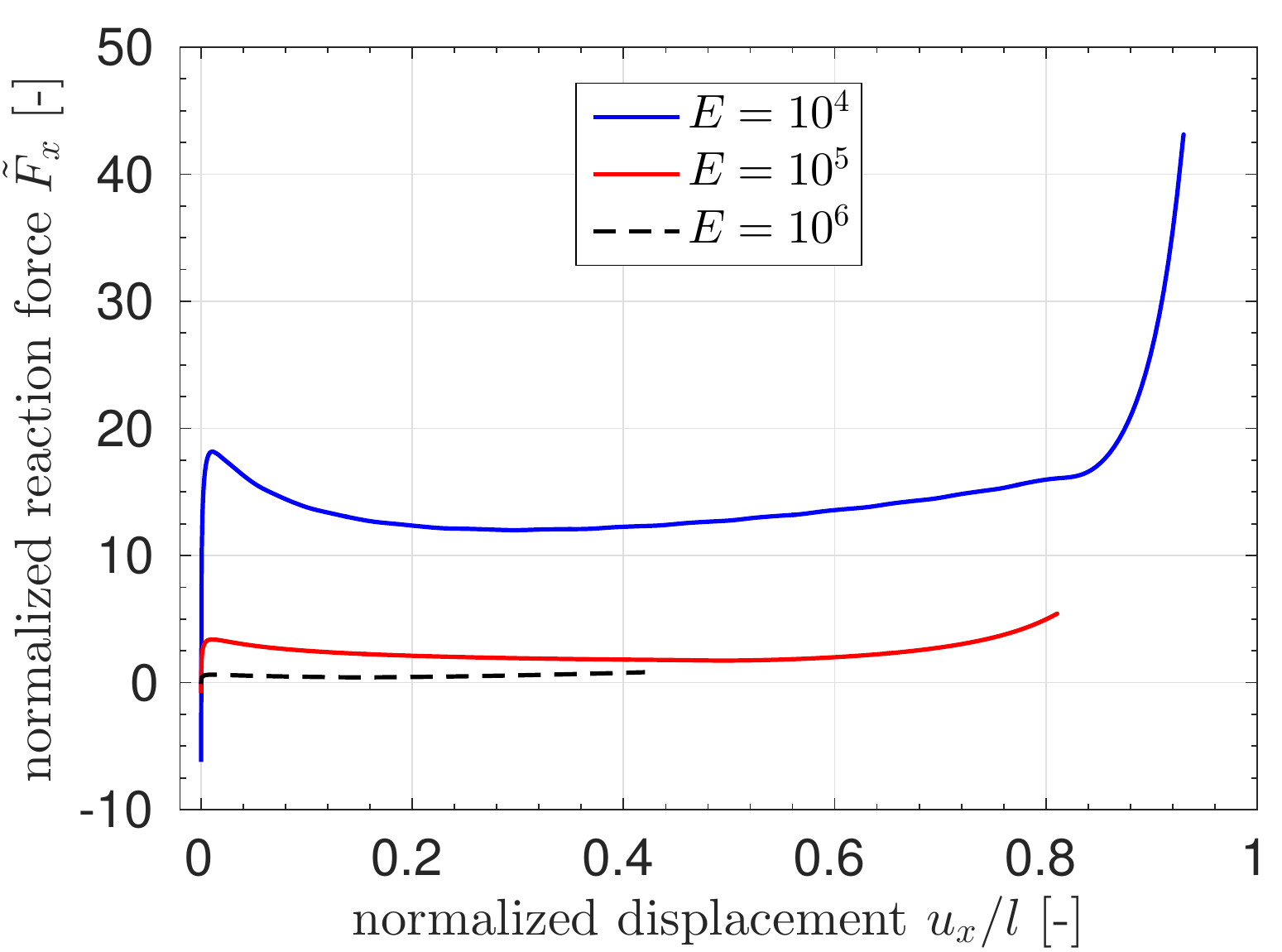}
    \label{fig::num_ex_elstat_attraction_twoparallelbeams_static_pulloff_reactionforce_over_displacement_comparison_youngsmodulus_alternativenormalization}
  }
  \caption{Comparison of results for ten times larger and ten times smaller value for Young's modulus~$E$.}
  \label{fig::num_ex_elstat_attraction_twoparallelbeams_static_pulloff_comparison_youngsmodulus}
\end{figure}
\begin{figure}[htpb]%
  \centering
  \subfigure[Static equilibrium configurations for~$u_x/l\approx0.4$.]{
    \vspace{-2cm}
    \hspace{-1cm}
    \includegraphics[width=0.32\textwidth]{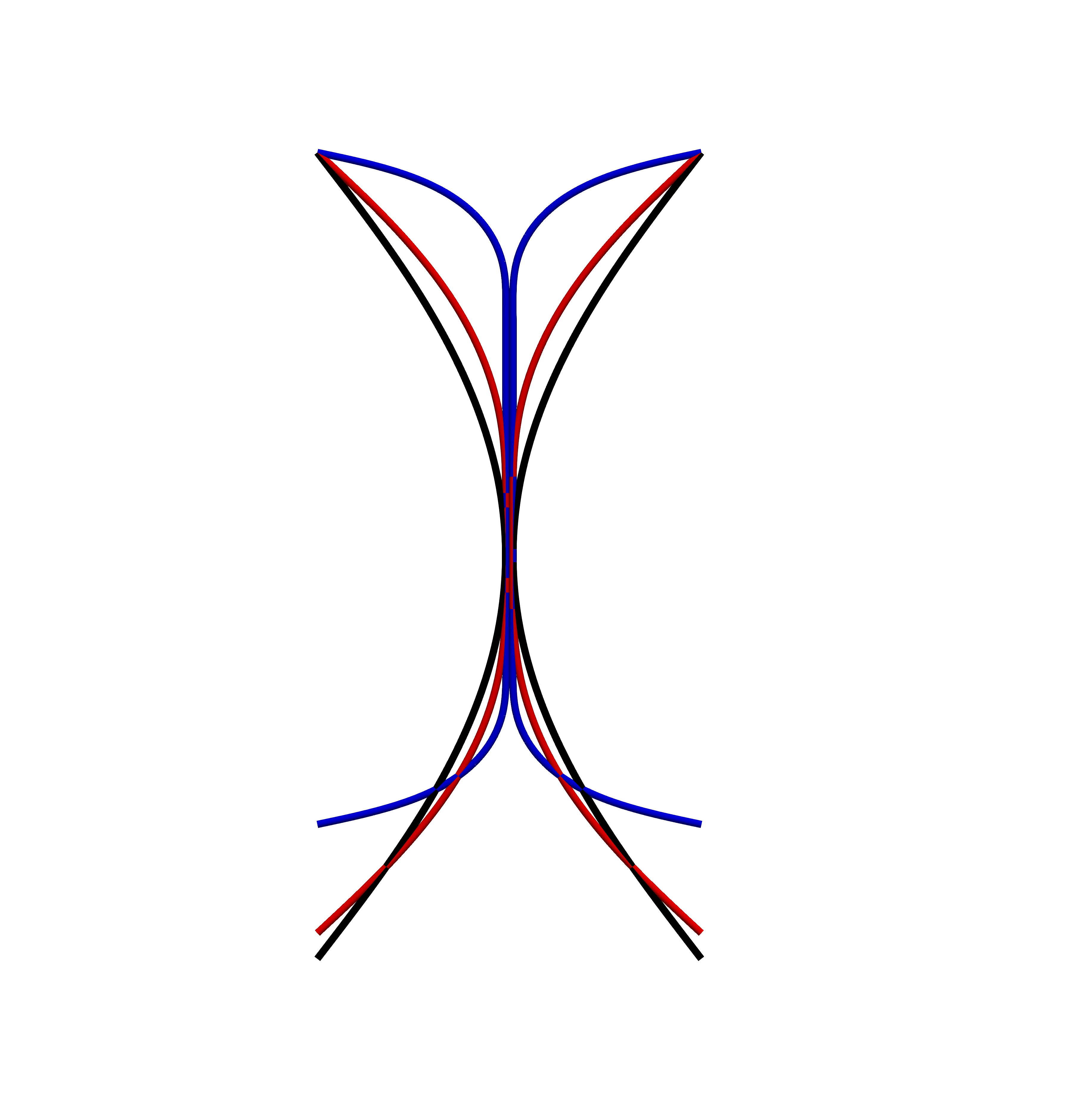}
    \hspace{-1cm}
    \label{fig::num_ex_elstat_attraction_twoparallelbeams_static_pulloff_comparison_youngsmodulus_normalizedseparation0_4}
  }
  \hspace{1cm}
  \subfigure[Static equilibrium configurations, ultimately before snapping free.]{
    \includegraphics[width=0.4\textwidth]{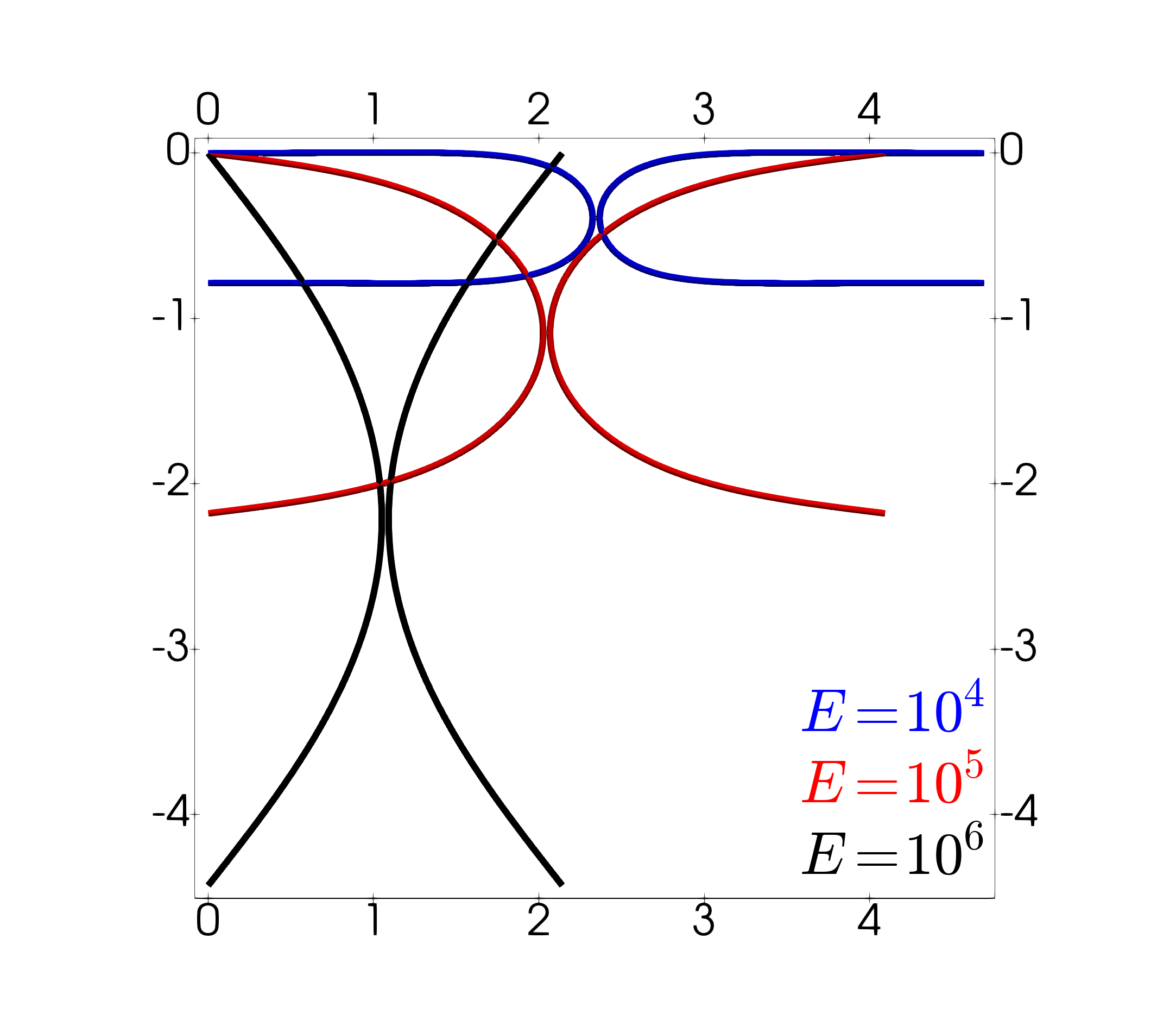}
    \label{fig::num_ex_elstat_attraction_twoparallelbeams_static_pulloff_comparison_youngsmodulus_finalstatebeforesnapfree}
  }
  \caption{Comparison of equilibrium configurations for ten times larger and ten times smaller value for Young's modulus~$E$.}
  \label{fig::num_ex_elstat_attraction_twoparallelbeams_static_pulloff_comparison_youngsmodulus_snapshots}
\end{figure}

The resulting force-displacement curves are shown in \figref{fig::num_ex_elstat_attraction_twoparallelbeams_static_pulloff_reactionforce_over_displacement_comparison_youngsmodulus}.
Accordingly, the static equilibrium configurations for one exemplarily chosen displacement value~$u_x/l=0.4$ are compared in~\figref{fig::num_ex_elstat_attraction_twoparallelbeams_static_pulloff_comparison_youngsmodulus_normalizedseparation0_4} and the configurations ultimately before snapping free, i.\,e., corresponding to the rightmost data point of each curve, are visualized in~\figref{fig::num_ex_elstat_attraction_twoparallelbeams_static_pulloff_comparison_youngsmodulus_finalstatebeforesnapfree}.
Note the two different variants of normalization of force values shown in~\figref{fig::num_ex_elstat_attraction_twoparallelbeams_static_pulloff_reactionforce_over_displacement_comparison_youngsmodulus} and~\figref{fig::num_ex_elstat_attraction_twoparallelbeams_static_pulloff_reactionforce_over_displacement_comparison_youngsmodulus_alternativenormalization}.
On the one hand, using the same reference force for all scenarios allows to compare the absolute force levels in~\figref{fig::num_ex_elstat_attraction_twoparallelbeams_static_pulloff_reactionforce_over_displacement_comparison_youngsmodulus}.
We find that the force peak associated with the initiation of fiber deformation is more pronounced for higher values of Young's modulus and thus bending stiffness of the fibers, as expected from the discussion of this initiation phase above.
Directly related to this, the subsequent peeling phase shows higher force values and passes over to the final pull-off phase at smaller displacement values.
The force plateau, which is characteristic for peeling with zero bending resistance, is not observable at all for the highest value of Young's modulus~$E=10^{6}$ considered here, but is very pronounced for~$E=10^{4}$.
In the final pull-off phase, fibers with higher Young's modulus again show higher force values, however with a less sharp increase just before snapping free.
Altogether, the system behavior shows some analogy to the failure of brittle and ductile material.

Alternatively, we can compute an individual reference force for each of the three curves with the corresponding value for the Young's modulus~$E=10^4, 10^5, 10^6$, respectively (see \figref{fig::num_ex_elstat_attraction_twoparallelbeams_static_pulloff_reactionforce_over_displacement_comparison_youngsmodulus_alternativenormalization}).
As expected, this reference force in the case of the ten times smaller / larger value of Young's modulus will be ten times smaller / larger than the reference force originally obtained for~$E=10^{5}$ and used in~\figref{fig::num_ex_elstat_attraction_twoparallelbeams_static_pulloff_reactionforce_over_displacement_comparison_youngsmodulus}, respectively.
This alternative normalization nicely illustrates the relative strength of the electrostatic interaction forces as compared to the forces that would typically result from large elastic bending deformation as represented by the scenario to compute the reference force.
We observe that the pull-off force in the case of the most flexible fibers with~$E=10^4$ exceeds this reference force by an impressive factor of more than~$40$.

Getting back to the comparison with the previous study of one-sided peeling of a thin film adhering to a rigid surface via vdW forces~\cite{Sauer2011c}, the results from variation of the relative adhesive strength over two orders of magnitude basically confirm the conclusions drawn above for the reference value.
The initial steep increase of the force and subsequent decrease in the initiation and peeling phase can be identified as unifying characteristics throughout all setups.
In addition, the trends of increasing force peak values and decreasing plateau width with increasing Young's modulus, i.\,e., decreasing relative strength of adhesion are observable both for the system here as well as for one-sided peeling of a thin film from a rigid substrate.
Also, the distinct pull-off phase with increasing force values including the global force maximum is reproduced here for all values of the relative strength of adhesion whereas it is not present for any of the parameter values in the one-sided peeling scenario of~\cite{Sauer2011c}.
This confirms the causal link between peeling from two sides and such a pronounced pull-off phase as suggested and explained above.

\begin{figure}[htpb]%
  \centering
  \subfigure[$E=10^4$]{
    \includegraphics[width=0.4\textwidth]{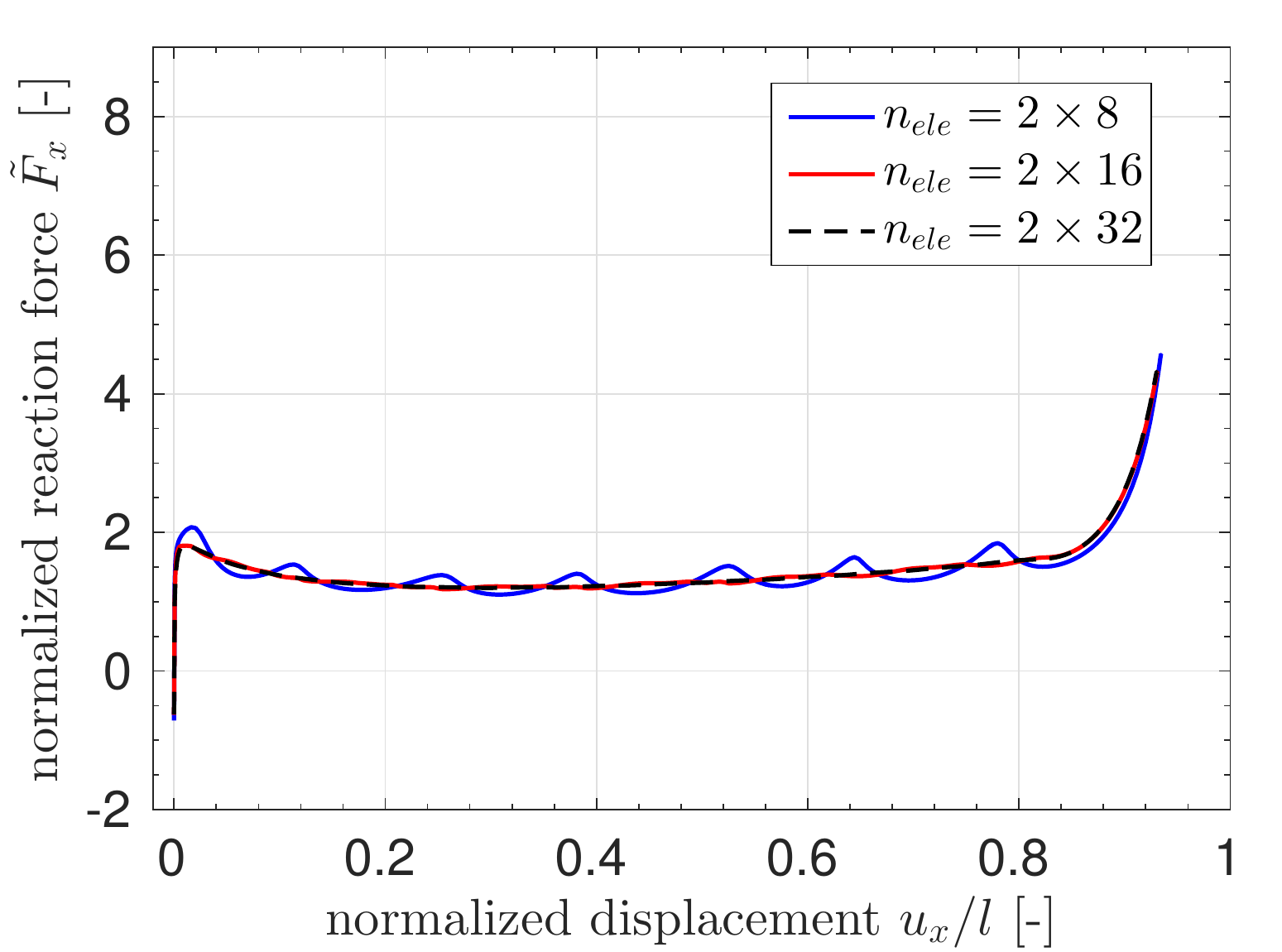}
    \label{fig::num_ex_elstat_attraction_twoparallelbeams_static_pulloff_mesh_refinement_young1e4}
  }
  \hspace{1cm}
  \subfigure[$E=10^5$]{
    \includegraphics[width=0.4\textwidth]{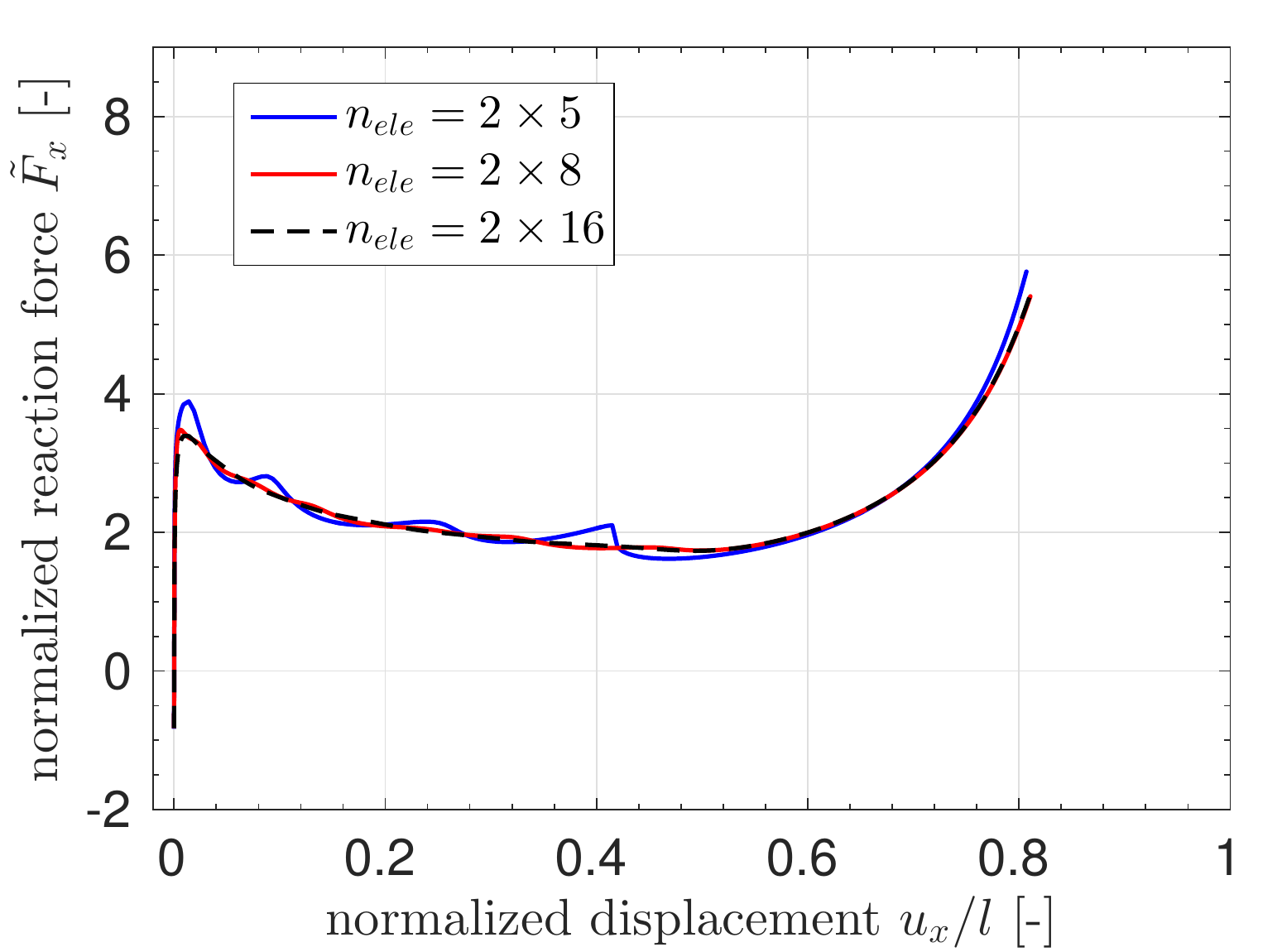}
    \label{fig::num_ex_elstat_attraction_twoparallelbeams_static_pulloff_mesh_refinement_young1e5}
  }
  \subfigure[$E=10^6$]{
    \includegraphics[width=0.4\textwidth]{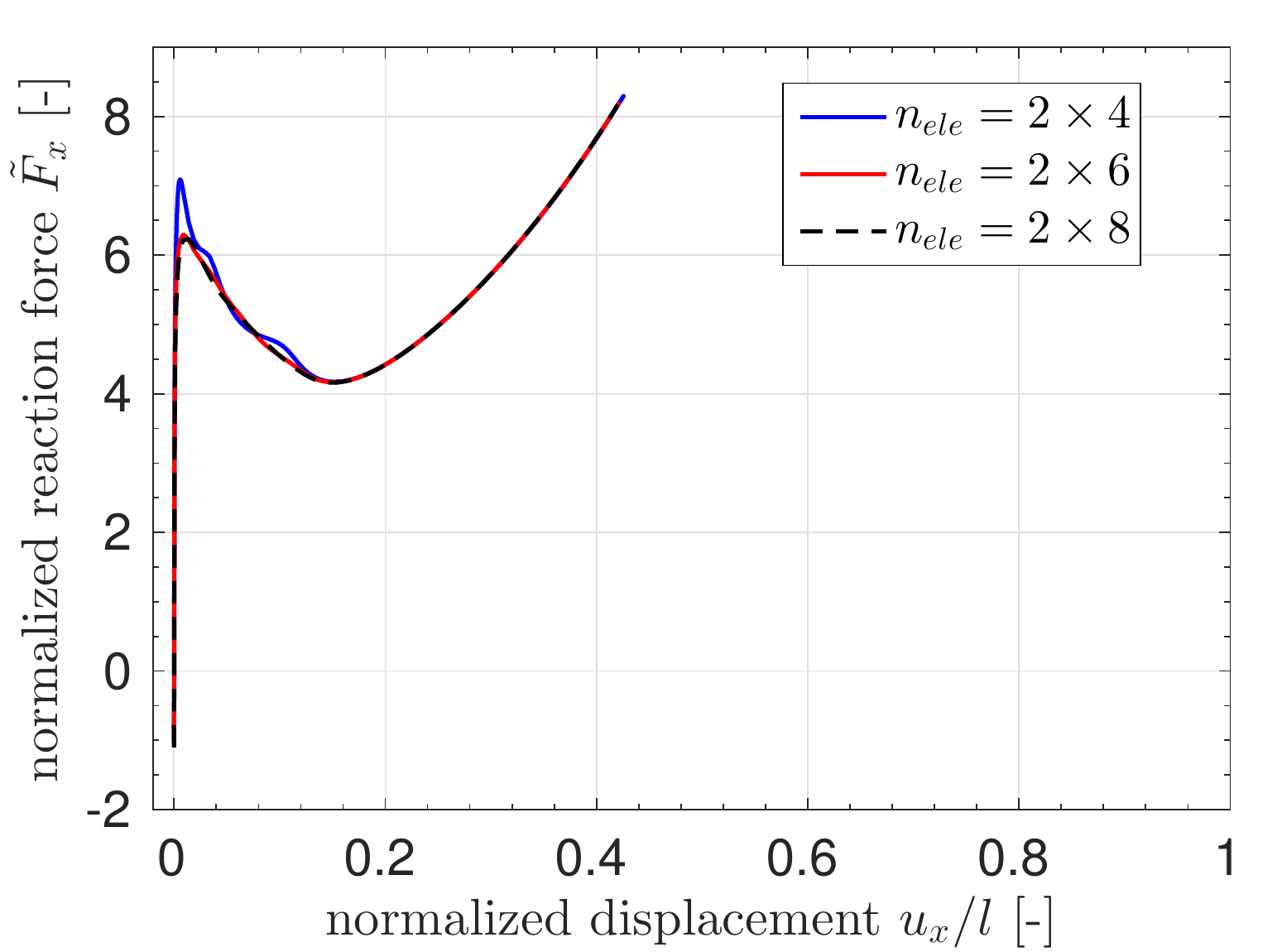}
    \label{fig::num_ex_elstat_attraction_twoparallelbeams_static_pulloff_mesh_refinement_young1e6}
  }
  \caption{Effect of mesh refinement depending on Young's modulus~$E$. Too coarse discretizations lead to artificial oscillations in the force-displacement curves. Two integration segments per element with ten Gauss points each, i.\,e.,~$n_\text{GP,tot,ele-length}=20$ are used in all simulations. However, doubling~$n_\text{GP,tot,ele-length}$ did not eliminate the oscillations for each of the three coarsest discretizations, such that the error from numerical quadrature is ruled out as the decisive factor. Here, we use the same normalization for the force values as in~\figref{fig::num_ex_elstat_attraction_twoparallelbeams_static_pulloff_reactionforce_over_displacement_comparison_youngsmodulus}.}
  \label{fig::num_ex_elstat_attraction_twoparallelbeams_static_pulloff_mesh_refinement}
\end{figure}

\paragraph{Discussion of the numerical approximation quality}$\,$\\
At the end of this section, we briefly analyze the important numerical aspects of spatial discretization as well as numerical integration error.
Generally, peeling simulations are known to be extremely sensitive to non-smoothness and coarseness of the spatial discretization and considerable effort has been made in the past to tackle this issue by surface enrichment strategies for 2D solid elements \cite{Sauer2011,Sauer2013a}.
As outlined in~\secref{sec::beam_theory}, here we use third order Hermite interpolation of the beam centerline, which directly ensures~$C^1$-continuity and needs no smoothing in the first place.
This carries over to a smooth representation of the inter-axis separation and hence the interaction force field.
We can thus apply relatively coarse meshes that are limited by the inherent spatial approximation error rather than non-smoothness at the element boundaries.
\figref{fig::num_ex_elstat_attraction_twoparallelbeams_static_pulloff_mesh_refinement} shows the results of our mesh refinement study.
For each of the values for Young's modulus used above, three different levels of mesh refinement are shown.
One of them obviously is too coarse and leads to artificial oscillations in the force values as typically observed in this case.
For the second and third discretization, the difference in results is already very small such that the second refinement level can arguably be regarded as a fine enough discretization for the purposes of this study.
To rule out the error from numerical quadrature as the decisive factor for the oscillations, we repeated each of the three simulations with the coarsest mesh and doubled the number of Gauss points per element from~$n_\text{GP,tot,ele-length}=20$ to $40$, which did not eliminate the oscillations.
From these results, it also becomes obvious that we need more elements for smaller values of the Young's modulus, i.\,e., more flexible fibers.
This can be explained by the degree of fiber deformation as visible e.\,g.~from \figref{fig::num_ex_elstat_attraction_twoparallelbeams_static_pulloff_comparison_youngsmodulus_finalstatebeforesnapfree}.
In order to limit the spatial discretization error to a minimum and ensure comparability of results, $32$ elements per fiber were used for all simulations in~\figref{fig::num_ex_elstat_attraction_twoparallelbeams_static_pulloff_comparison_youngsmodulus}.
We can thus conclude that the smooth, cubic centerline representation used in combination with the SSIP approach throughout this work allows for robust and accurate peeling simulations even with relatively coarse meshes.

\section{Van der Waals attraction}\label{sec::num_ex_vdW_attraction_twoparallelbeams_pulloff_from_contact}
The aim of this section is to repeat the peeling experiment of~\secref{sec::num_ex_elstat_attraction_twoparallelbeams_peeling_pulloff} with a fundamentally different type of attractive forces, namely vdW forces.
This allows us to analyze the differences and similarities in the force response of the system and also to discuss the differences in the numerical methods used to model the two phenomena.

\subsection{Setup and parameters}
\begin{figure}[htpb]%
  \centering
  \subfigure[Problem setup: undeformed configuration.]{
    \def\svgwidth{0.13\textwidth}
\begingroup%
  \makeatletter%
  \providecommand\color[2][]{%
    \errmessage{(Inkscape) Color is used for the text in Inkscape, but the package 'color.sty' is not loaded}%
    \renewcommand\color[2][]{}%
  }%
  \providecommand\transparent[1]{%
    \errmessage{(Inkscape) Transparency is used (non-zero) for the text in Inkscape, but the package 'transparent.sty' is not loaded}%
    \renewcommand\transparent[1]{}%
  }%
  \providecommand\rotatebox[2]{#2}%
  \newcommand*\fsize{\dimexpr\f@size pt\relax}%
  \newcommand*\lineheight[1]{\fontsize{\fsize}{#1\fsize}\selectfont}%
  \ifx\svgwidth\undefined%
    \setlength{\unitlength}{90bp}%
    \ifx\svgscale\undefined%
      \relax%
    \else%
      \setlength{\unitlength}{\unitlength * \real{\svgscale}}%
    \fi%
  \else%
    \setlength{\unitlength}{\svgwidth}%
  \fi%
  \global\let\svgwidth\undefined%
  \global\let\svgscale\undefined%
  \makeatother%
  \begin{picture}(1,2.5)%
    \lineheight{1}%
    \setlength\tabcolsep{0pt}%
    \put(0,0){\includegraphics[width=\unitlength,page=1]{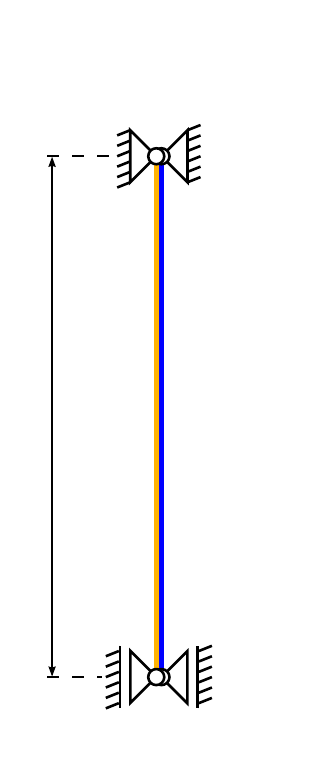}}%
    \put(0.72220523,0.16035767){\color[rgb]{0,0,0}\makebox(0,0)[lt]{\lineheight{0}\smash{\begin{tabular}[t]{l}$u_x, F_x^{br}$\end{tabular}}}}%
    \put(0.19677776,1.13176587){\color[rgb]{0,0,0}\makebox(0,0)[lt]{\lineheight{0}\smash{\begin{tabular}[t]{l}$l$\end{tabular}}}}%
    \put(0.71669108,2.17380371){\color[rgb]{0,0,0}\makebox(0,0)[lt]{\lineheight{0}\smash{\begin{tabular}[t]{l}$u_x, F_x^{tr}$\end{tabular}}}}%
    \put(0,0){\includegraphics[width=\unitlength,page=2]{num_ex_vdW_twoparallelbeams_pulloff_from_contact_problem_setup.pdf}}%
    \put(0.54009603,2.26407878){\color[rgb]{0,0,0}\makebox(0,0)[lt]{\lineheight{0}\smash{\begin{tabular}[t]{l}$y$\end{tabular}}}}%
    \put(0.81509603,2.03602702){\color[rgb]{0,0,0}\makebox(0,0)[lt]{\lineheight{0}\smash{\begin{tabular}[t]{l}$x$\end{tabular}}}}%
    \put(0,0){\includegraphics[width=\unitlength,page=3]{num_ex_vdW_twoparallelbeams_pulloff_from_contact_problem_setup.pdf}}%
  \end{picture}%
\endgroup%

    \label{fig::num_ex_vdW_twoparallelbeams_problem_setup}
  }
  \hspace{0.5cm}
  \subfigure[Quasi-static force-displacement curve. Force values to be interpreted as multiple of a reference point load that causes a deflection of~$l/4$ if applied at the fiber midpoint.]{
    \includegraphics[width=0.45\textwidth]{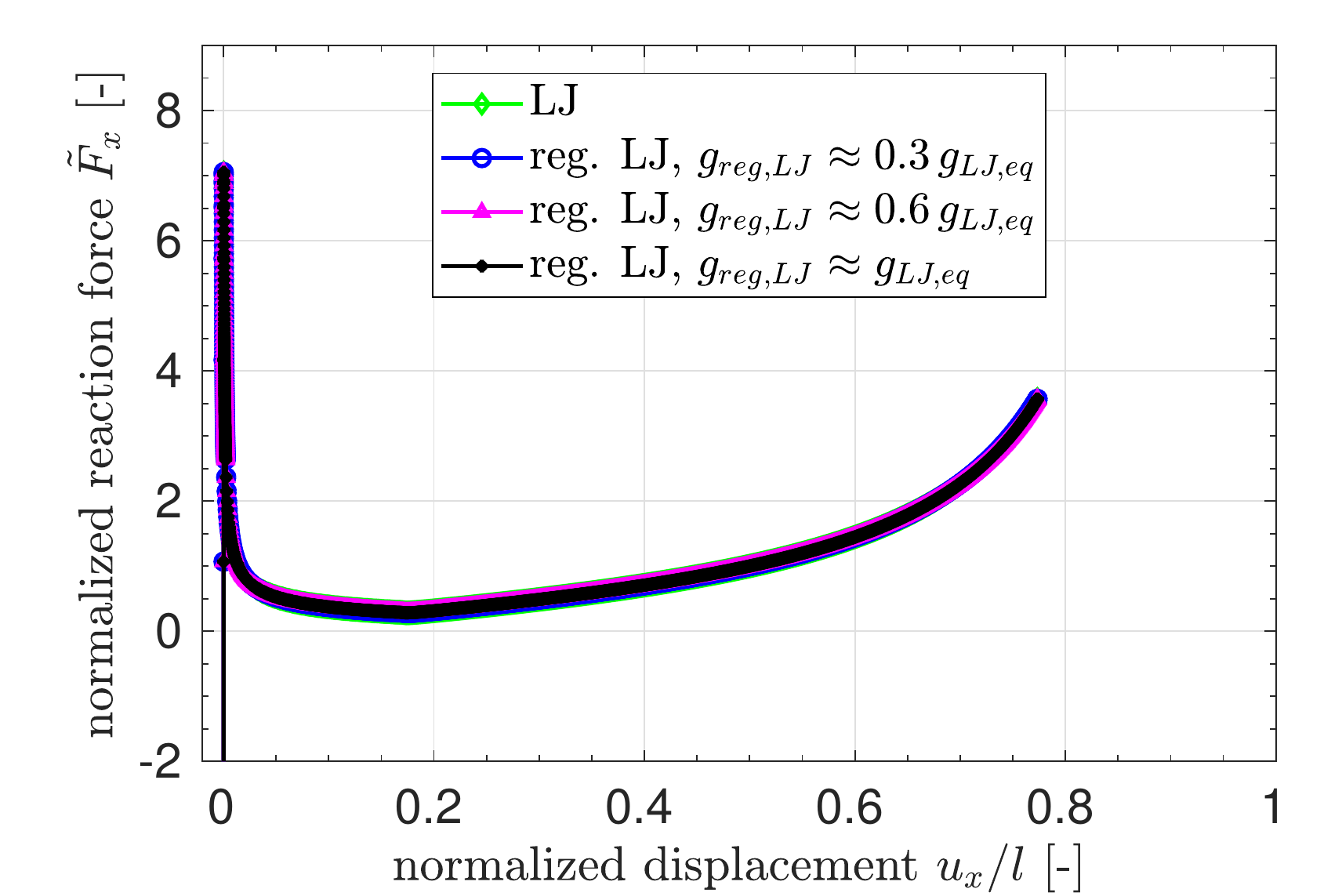}
    \label{fig::num_ex_vdW_attraction_twoparallelbeams_pulloff_force_over_displacement}
  }
  \hfill
  \subfigure[Detail view for small separations.]{
    \includegraphics[width=0.3\textwidth]{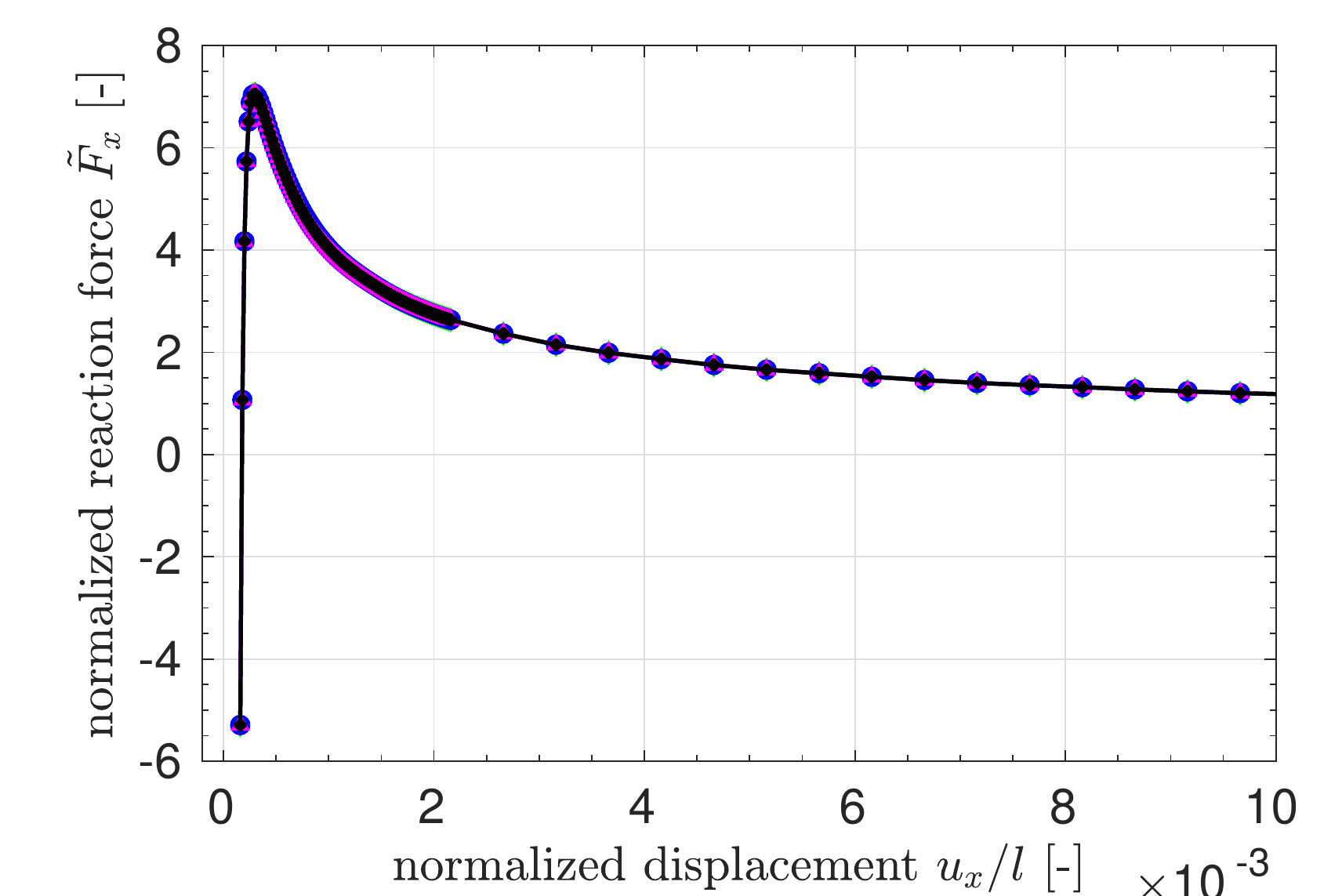}
    \label{fig::num_ex_vdW_attraction_twoparallelbeams_pulloff_force_over_displacement_zoom}
  }
  \caption{Numerical peeling experiment with two adhesive elastic fibers interacting via the LJ potential.}
  \label{fig::num_ex_vdW_attraction_twoparallelbeams_static}
\end{figure}

As mentioned above, the setup of this numerical experiment shown in~\figref{fig::num_ex_vdW_twoparallelbeams_problem_setup} is almost identical to the one discussed in the preceding~\secref{sec::num_ex_elstat_attraction_twoparallelbeams_peeling_pulloff}.
The most important difference is the fact that the fibers interact via a LJ potential instead of the electrostatic one.
Accordingly, the LJ interaction is evaluated using the SSIP approach as outlined in~\secref{sec::SSIP_vdW} and~\secref{sec::SSIP_repLJ} for the attractive vdW and repulsive part, respectively.
The parameter values of the LJ point pair potential law are chosen exemplarily as~$k_\text{vdW}=-10^{-7}$ and~$k_\text{repLJ}=5\times10^{-25}$ such that the resulting reaction forces are of the same order of magnitude as for the electrostatic adhesion.
Obviously, this is an arbitrary choice and the comparison of force-displacement curves between electrostatic and LJ interaction later on will only be a qualitative one.
The particle densities are assumed to be constant and set to~$\rho_{1/2}=1.0$.
Due to the rapid decay of both the adhesive vdW and even more the repulsive part of the LJ potential, the interaction has an extremely short range and it is essential to ensure a fine resolution of these small length scales.
Thus, a large number of~$64$ elements per fiber and five integration segments with ten Gauss points each is used for the discretization and numerical integration of the contributions from LJ interaction and we verified that a further refinement does not change the results perceptibly.
To reduce the computational cost, the very short range of the interactions has been exploited by using a cut-off radius~$r_c=0.1=5R$ which again did not change the results perceptibly.
See~\secref{sec::num_ex_elstat_attraction_twoparallelbeams_peeling_pulloff_setup} for all geometric and material parameter values.
Specifically, we again use the original value~$E=10^5$ for the Young's modulus, which has been varied in the parameter study at the end of the preceding section.
The one, yet important difference in the geometric setup of the problem is the initial separation of the fibers in the first step of the simulation that turns out to be crucial in order to be close enough to a static equilibrium configuration such that the nonlinear solver converges.
For this reason, an analytical solution for the equilibrium separation~$g_\text{LJ,eq,cyl$\parallel$cyl}$ of two parallel, infinitely long cylinders interacting via the LJ potential has been derived in~\cite{GrillSSIP} and is repeated here for convenience:
\begin{equation}\label{eq::equilibrium_spacing_LJ_cylinders_parallel_smallsep}
  g_\text{LJ,eq,cyl$\parallel$cyl} \approx \num{0.57169} \, r_\text{LJ,eq}
\end{equation}
Here, $r_\text{LJ,eq}$ denotes the equilibrium spacing of a point pair interacting via the LJ potential, which is related to the alternative set of parameters used in this work according to~$r_\text{LJ,eq} = \left( - 2 \, k_\text{repLJ} / k_\text{vdW}  \right)^{1/6}$.
Refer to~\cite{GrillSSIP} for the required derivation of the cylinder-cylinder LJ interaction potential as well as force law starting from the point-point LJ potential law.
For the parameters listed above, we obtain~$g_\text{LJ,eq,cyl$\parallel$cyl} \approx 4.2 \times 10^{-2} \cdot R \approx 1.7 \times 10^{-4} \cdot l$.
In order to include also the repulsive regime, a slightly smaller initial separation, i.\,e., displacement~$u_x/l = 1.6 \times 10^{-4}$ is chosen here.
Note however, that in contrast to infinitely long cylinders considered in the theory of eq.~\eqref{eq::equilibrium_spacing_LJ_cylinders_parallel_smallsep}, the force-free equilibrium configuration for this pair of deformable fibers with finite length~$l$ is not straightforward to find.
In particular, it is not the trivial case of two straight fibers at a constant surface-to-surface spacing~$g$ along their length.

\subsection{Results and discussion}\label{sec::results_num_ex_vdW_attraction_twoparallelbeams_pulloff_from_contact}
\begin{figure}[htpb]%
  \centering
    \includegraphics[width=0.45\textwidth]{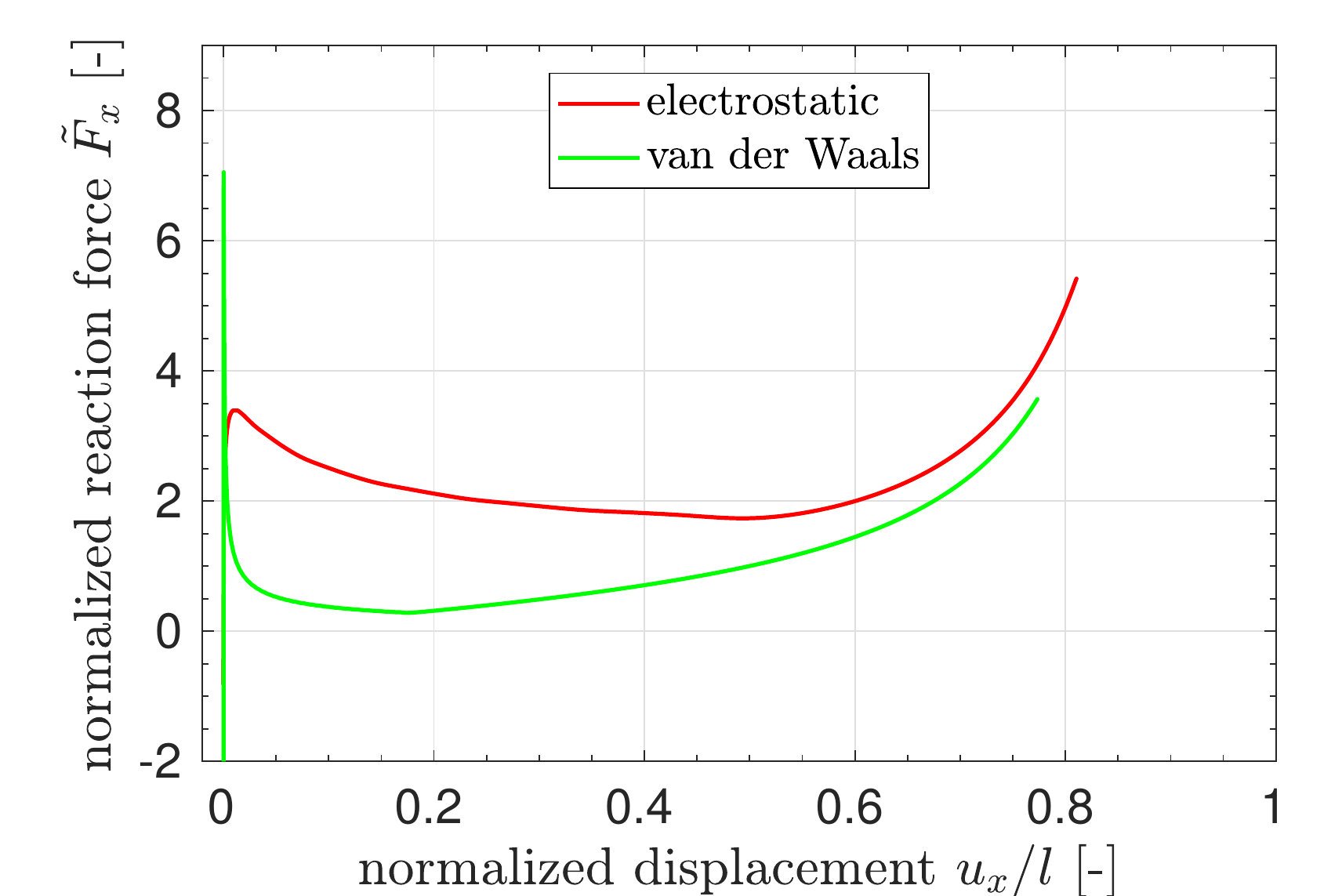}
  \caption{Comparison of the quasi-static force-displacement curves obtained for electrostatic and vdW attraction. Force values to be interpreted as multiple of a reference point load that causes a deflection of~$l/4$ if applied at the fiber midpoint.}
  \label{fig::num_ex_vdW_attraction_twoparallelbeams_static_pulloff_vdW_vs_elstat_force_over_displacement}
\end{figure}

\figref{fig::num_ex_vdW_attraction_twoparallelbeams_pulloff_force_over_displacement} finally shows the resulting force-displacement curve.
In addition, the most interesting range of very small displacement values~$u_x/l<10^{-2}$ is magnified and shown in a separate plot in~\figref{fig::num_ex_vdW_attraction_twoparallelbeams_pulloff_force_over_displacement_zoom}.
Let us first leave the different variants of numerical regularization aside, since all of them yield identical results and will be discussed later.
As suggested by the analytical solution for the equilibrium spacing of infinitely long, parallel cylinders, the first data point with the slightly smaller initial separation lies in the repulsive regime with~$\tilde{F}_x<0$, whereas all subsequent data points yield positive, i.\,e., tensile reaction forces.
The qualitative comparison with the electrostatic attraction shown in~\figref{fig::num_ex_vdW_attraction_twoparallelbeams_static_pulloff_vdW_vs_elstat_force_over_displacement} reveals a substantial difference in the system response, most obvious in terms of the much sharper and also higher force peak during the initiation of the peeling process.
Interestingly, after quickly dropping to a much smaller value, the adhesive vdW force effectively keeps the fibers in contact up to a comparable separation of the fibers' endpoints as in the electrostatic case (see again~\figref{fig::num_ex_elstat_attraction_twoparallelbeams_pulloff_separation4_00_incontact} for a visualization of the corresponding system state).

At this point, recall the known limitation of the applied SSIP law for vdW adhesion outlined in~\secref{sec::SSIP_vdW}.
To ensure that the model error does not alter the qualitative results nor the following conclusions, we have repeated the simulation employing a prototype of a novel computational model with correct asymptotic scaling behavior and thus significantly enhanced accuracy.
This confirmed that the strength of adhesion resulting from the given value of~$k_\text{vdW}$ is overestimated here (which is expected also from the analysis in~\cite{GrillSSIP}), resulting in a higher force peak and larger pull-off displacement, however the characteristic shape of the force-displacement curve and thus all the following conclusions remain valid.
In fact, it can be shown that the prefactor~$k_\text{vdW}$ of the simple SSIP law can be calibrated in a manner such that the resulting overall force-displacement curve is in excellent agreement with the corresponding force-displacement curve resulting from the enhanced computational model.
In this context, note also that the applied set of parameter values~$k_\text{vdW}, k_\text{repLJ}$ for the LJ interaction is chosen based on the heuristic criterion that the resulting reaction forces are of the same order of magnitude as in the case of electrostatic adhesion considered in~\secref{sec::num_ex_elstat_attraction_twoparallelbeams_peeling_pulloff}, such that this comparison between both cases naturally is a qualitative rather than quantitative one.

Following up on the analyses of the previous sections, we can conclude that the characteristic shape of the force-displacement curve in the initiation and peeling phases is obtained here as well, although the force peak is higher, sharper and shifted to smaller displacement values, which is most likely a result of the shorter range of adhesive forces.
Once again, also the pronounced pull-off phase is observed (despite the shorter range of interaction), which supports the argument that it can be attributed to the symmetric two-sided peeling from both fiber endpoints.
As a final and most important result, however, it has to be pointed out that the maximum force value in the entire separation process occurs during initiation of the peeling in the case of vdW adhesion considered here, whereas it occurs in the final pull-off phase ultimately before snapping free in the case of electrostatic attraction studied in~\secref{sec::num_ex_elstat_attraction_twoparallelbeams_peeling_pulloff}.
This noteworthy finding can again be explained by the fundamental difference between the short and long range of these interaction types.

\paragraph{Discussion of the numerical regularization of the LJ interaction law}$\,$\\
A suitable regularization strategy to remedy the inherent singularity of the LJ SSIP law has been proposed in~\cite{GrillSSIP}.
The concrete numerical example of this section allows us to study and quantify its significant positive impact on the performance of the nonlinear solver and in this way complement the theoretical considerations made in~\cite{GrillSSIP}.
The LJ potential applied here shows both very high gradients in the force-distance law as well as the singularity in the SSIP law for zero separation~$g=0$ and is thus the more challenging case with respect to the nonlinear solver as compared to electrostatic interaction.
In order to find a solution of the nonlinear system of equations in every load step, we apply Newton's method in combination with a displacement increment control as outlined in~\secref{sec::methods} (see~\cite{GrillSSIP} for details).
Using a very strict upper bound for the displacement increment per iteration of~$|\Delta u|_\text{max}= R/20$, we were able to compute the solution for the LJ interaction without any modification of the SSIP laws stated in~\secref{sec::SSIP_vdW} and~\ref{sec::SSIP_repLJ} for the vdW and repulsive part, respectively.
This effectively avoids the singularity in any unconverged state, however comes at the tremendous computational cost of an average of~$46.2$ required Newton iterations in each of the approximately~$1600$ steps required to compute the entire force-displacement curve shown in~\figref{fig::num_ex_vdW_attraction_twoparallelbeams_pulloff_force_over_displacement} (green line with diamonds).
These numbers underline the urgent need for the regularization of the LJ SSIP law in the limit of zero separation~$g\to0$ as proposed in~\cite{GrillSSIP}.
Applying the proposed linear extrapolation of the section-section interaction force law below a certain separation~$g_\text{reg,LJ}$ indeed considerably improves the performance of the nonlinear solver.
The average number of Newton iterations per step decreases to~$10.2$, which is almost a factor of five, while the results shown in~\figref{fig::num_ex_vdW_attraction_twoparallelbeams_pulloff_force_over_displacement} (blue, pink, and black line) coincide with the full LJ solution (green line) down to machine precision.
This remarkable reduction of computational cost while obtaining identical results is exactly the same for all three values of the regularization parameter~$g_\text{reg,LJ}= \{0.3,0.6,1.0\} \times g_\text{LJ,eq,cyl$\parallel$cyl}$ that we applied.
As outlined in the theoretical considerations of~\cite{GrillSSIP}, this is reasonable and expected, because we choose a regularization parameter~$g_\text{reg,LJ} \leq g_\text{LJ,eq,cyl$\parallel$cyl}$ that is smaller than (or equal to) any separation value~$g$ occurring anywhere in the system in any converged state.
Thus, the solution never ``sees'' the modification of the LJ force law in the interval~$g<g_\text{reg,LJ}$ and the results are identical.
If, on the contrary, the regularization parameter is chosen as~$g_\text{reg,LJ}=1.2 \times g_\text{LJ,eq,cyl$\parallel$cyl} > g_\text{LJ,eq,cyl$\parallel$cyl}$, we observed that the nonlinear solver failed to converge at~$u_x/l \approx 0.17$ and the obtained force values in the range~$u_x/l \lessapprox 0.17$ deviate from those for the unmodified LJ SSIP law.
This underlines the importance of the correct choice of the regularization parameter and the knowledge of the theoretical equilibrium spacing~$g_\text{LJ,eq,cyl$\parallel$cyl}$ stated in eq.~\eqref{eq::equilibrium_spacing_LJ_cylinders_parallel_smallsep}.

\section{Summary, conclusions and outlook}\label{sec::conclusion_outlook}
This article studies the fundamental problem of separating two adhesive elastic fibers based on numerical simulation employing a finite element model for molecular interactions between curved slender fibers, which has recently been developed by the authors~\cite{GrillSSIP}.
Specifically, it covers the peeling and pull-off process starting from fibers contacting along its entire length to fully separated fibers (and also the reverse order) including all intermediate configurations and the well-known physical instability of snapping into contact and snapping free.
In order to study the key influences, the strength of adhesion relative to the Young's modulus of the fibers has been varied over a broad range of values spanning two orders of magnitude, and also two different types of attractive forces resulting either from van der Waals (vdW) interactions or the electrostatic interaction of oppositely charged non-conducting fibers are considered.
We have analyzed the resulting force-displacement curve revealing a rich, highly nonlinear system behavior and thoroughly investigated the underlying physical mechanisms arising from the interplay of adhesion, mechanical contact interaction and structural resistance against (axial, shear and bending) deformation.
Based on the differences in these fundamental mechanisms, the three distinct phases of a) initiation of fiber deformation and peeling, b) the actual peeling, and c) a final pull-off phase have been identified.
The initiation phase is characterized by a steep initial slope towards a sharp force peak and followed by the peeling phase with gradually decreasing force values eventually approaching a plateau-like regime of almost constant peeling force.
The unitary nature of these first two phases in the peeling of adhesive elastic structures is confirmed by the comparison with previous studies and across all considered variants in this study.
On the contrary, the presence of the pull-off stage as a third distinct phase of the separation process, that is characterized by a significant increase of the force over an extended range of displacement values before finally snapping free, was not observed in the aforementioned one-sided peeling studies and can thus be attributed to the application of pulling forces at both ends of the fibers, which results in a two-sided instead of a one-sided peeling.
Moreover, the practically highly relevant global maximum of the pulling force is found to occur at the end of the initiation phase in case of short-ranged vdW attraction, whereas it occurs in the final pull-off phase ultimately before snapping free in the case of the long-ranged electrostatic attraction.

The complexity of the physical effects of adhesion, contact and elasticity in the regime of large deformations -- individually and particularly in combination -- carries over to the computational models and numerical solution methods required for this study.
In addition to the physical behavior of the system, we have therefore discussed the decisive aspects regarding robustness and accuracy of the simulations.
Here, the major challenges include the delicate task of determining equilibrium configurations in the direct vicinity of the mentioned physical instability, the control of spatial discretization and numerical integration error such that the high gradients of short-range interaction potential laws are represented with sufficient accuracy as well as the regularization of these inverse power laws to remedy the singularity at zero separation.
Concretely, the regularization procedure for the employed beam-beam interaction model, which has been proposed in~\cite{GrillSSIP}, has proven to significantly enhance robustness and efficiency at identical accuracy by saving a factor of five in the number of nonlinear iterations when applied to the highly challenging example of Lennard-Jones interaction.
Besides the insights gained into the peeling and pull-off behavior of the specific two-fiber system, the present work thus serves as a proof of concept facilitating future applications of the employed model to increasingly complex systems of slender fibers.

Subsequent studies may include e.\,g.~the dynamics of the peeling and pull-off process or further scenarios of loading and support of the fibers, e.\,g.~twisting and out-of-plane bending, all of which is directly accessible by means of the computational models and methods applied in this work.
In the wider context of the authors' continued work on (the computational study of) fibrous biophysical systems on the microscale~\cite{mueller2014,Mueller2015,Slepukhin2019}, investigating the influence of adhesion on the self-assembly and mechanical behavior of biological fibrillar assemblies such as collagen or muscle fibers is considered a highly relevant subject of future research.
Given the importance of charge screening effects in aqueous electrolyte solution, an extension of the employed beam-beam interaction model in this respect would be a promising next step from a modeling point of view.

\appendix

\bibliography{library_PeelingPulloffAdhesiveFibers.bib}

\end{document}